\documentclass[10pt,journal]{IEEEtran}
\usepackage{algorithm}
\usepackage{algpseudocode}
\usepackage{amsfonts}
\usepackage{bm}
\usepackage{amsmath}
\usepackage{cite}
\usepackage{float}
\usepackage{amssymb}
\usepackage{enumerate}
\usepackage{color,xcolor}
\usepackage{graphicx}
\usepackage{multirow,tabularx}
\usepackage{subfigure}

\usepackage{cases}

\usepackage{xspace}
\usepackage{fixfoot}

\DeclareFixedFootnote*{\fp}{ Two different CRLBs for MSE depends on the amount of knowledge of the estimators about the sparsity structure of
the targeted deterministic signal \cite{C_Carbonelli_sparse}. Specifically, compared with the LSE, our proposed algorithm can estimate the sparse information (positions of non-zero entries) by employing the SMP and EM. Therefore, CRLBs are different for the LSE and LSE-SMP algorithm, which are denoted as $\mathbf{CRLB_{LSE}}$ and $\mathbf{CRLB_{LSE-SMP}}$ respectively.}

\makeatletter
\renewcommand\normalsize{%
   \@setfontsize\normalsize\@xpt\@xiipt
   \abovedisplayskip 1\p@ \@plus2\p@ \@minus3\p@
   \abovedisplayshortskip 5\z@ \@plus2\p@ \@minus3\p@
   \belowdisplayshortskip 5\p@ \@plus2\p@ \@minus3\p@
   \belowdisplayskip \abovedisplayskip
   \let\@listi\@listI}
\makeatother

\ifCLASSINFOpdf

\else

\fi

\begin{document}
%
\title{ Iterative Channel Estimation Using LSE and Sparse Message Passing for MmWave MIMO Systems }

\author{\IEEEauthorblockN{Chongwen Huang, \emph{Student Member, IEEE,} Lei Liu, \emph{Student Member, IEEE,} \\ Chau Yuen, \emph{Senior Member, IEEE,} Sumei Sun, \emph{Fellow, IEEE} }
\thanks{  The work of C. Yuen was supported by the MIT-SUTD International design center and NSFC 61750110529 Grant, and that of C. Huang by the PHC Merlion PhD program.

Chongwen Huang and Chau Yuen are with the Singapore Unversity of Technology and Design, Singapore. Lei Liu is with the Singapore Unversity of Technology and Design and City University of Hong Kong, Hong Kong, China. Sumei Sun is with the Institute for Infocomm Research (I2R), Agency for Science, Technology and Research (A$^{\star}$STAR), 138632, Singapore. (e-mail: chongwen$\_$huang@mymail.sutd.edu.sg, leiliuxidian@gmail.com, yuenchau@sutd.edu.sg, sunsm@i2r.a-star.edu.sg).
}
\thanks{ The material in this paper was presented in part at the conference of the IEEE Globecom 2016 workshop on mobile communications in higher frequency bands, Washington D.C., USA, Dec. 2016  \cite{Chongwen_01}.}
}
\maketitle
\begin{abstract}
We propose an iterative channel estimation algorithm based on the Least Square Estimation (LSE) and Sparse Message Passing (SMP) algorithm for the Millimeter Wave (mmWave) MIMO systems. The channel coefficients of the mmWave MIMO are approximately modeled as a Bernoulli-Gaussian distribution and the channel matrix is sparse with only a few non-zero entries. By leveraging the advantage of sparseness, we propose an algorithm that iteratively detects the exact locations and values of non-zero entries of the sparse channel matrix. At each iteration, the locations are detected by the SMP, and values are estimated with the LSE. We also analyze the Cram\'{e}r-Rao Lower Bound (CLRB), and show that the proposed algorithm is a minimum variance unbiased estimator under the assumption that we have the partial priori knowledge of the channel. Furthermore, we employ the Gaussian approximation for message densities under density evolution to simplify the analysis of the algorithm, which provides a simple method to predict the performance of the proposed algorithm. Numerical experiments show that the proposed algorithm has much better performance than the existing sparse estimators, especially when the channel is sparse. In addition, our proposed algorithm converges to the CRLB of the genie-aided estimation of sparse channels with only five turbo iterations.
\end{abstract}
\begin{IEEEkeywords}
Millimeter wave, iterative channel estimation, sparse message passing, Gaussian-Bernoulli distribution, Cram\'{e}r-Rao lower bound, minimum variance unbiased estimator, Gaussian approximation.
\end{IEEEkeywords}

%
\IEEEpeerreviewmaketitle
\section{Introduction}
Millimeter wave (mmWave) has been receiving tremendous interests from the academia, industry, and government for future 5G cellular systems \cite{Chongwen_01,rappaport_mmWave_model2013,mmWave_Channel_Modeling_evaluation,Dai_linglong_02,hcw_ICASSP_2018,rappaport_book_mmWave_2014,SP_for_2016,veug01,veug02,veug03} due to the available spectrum from 30 GHz to 300 GHz. However, mmWave poses new challenges. One of them is the severe path-loss. Recent urban model experiments show that path losses are 40 dB worse at 28 GHz compared to 2.8 GHz\cite{rappaport_mmWave_model2015,rappaport_mmWave_model2016}.

One way to overcome this severe path-loss of mmWave signal propagation is to increase the number of transmit and receive antennas \cite{SP_for_2016,xiao_ming,O_El_Ayach}. With the large number of antennas and relatively fewer channel paths, the mmWave channel is approximately sparse \cite{SP_for_2016,Sayeed_SP2002,P_Schniter_virtual,O_El_Ayach,P_Schniter_virtual_sparse}. The sparse feature is recently verified by measurements, for example, \cite{rappaport_mmWave_model2015,rappaport_mmWave_model2016,A.Ghosh_LOS} reported that mmWave channels typically exhibit only 3-4 scattering clusters in dense-urban non-line-of-sight environments. Therefore, conventional MIMO iterative channel estimation methods \cite{factor_graph_LDPC,Flanagan_iterative_estimation,qh_Guo_message,Y_Zhu_MP,JW_Choi_estimation,wusheng_MP,gqh_2015} are not suitable for mmWave systems due to the different channel characteristic and system model (i.e., mmWave systems usually employ the hybird anlog/digital architecture for reducing hardware cost and power consumption \cite{rappaport_mmWave_model2015,SP_for_2016}.). This prompts the need to design efficient channel estimation techniques for the mmWave systems.

For sparse channel estimation, several algorithms have been proposed in \cite{LASSO,Donoho_AMP,OMP01,T_SBL,C_Carbonelli_sparse,P_Schniter_EM,Alk_channel_Estimation,Guo_Qing}. They can be classified into three categories according to the required priori information (except the noise variance) of the channel. The algorithm in the first category requires to know the full knowledge of the channel's distribution, structure, etc., for example, the approximate message passing (AMP) algorithm \cite{Donoho_AMP} proposed by Donoho and Maleki. AMP is a low-complexity iterative Bayesian algorithm that can achieve approximately maximum a posteriori and minimum mean-squared error signal estimates. Iterative Detection/Estimation With Threshold (ITD-SE) \cite{C_Carbonelli_sparse}, Adaptive Compressed Sensing (ACS) estimation algorithm proposed in \cite{Alk_channel_Estimation} and Orthogonal Matching Pursuit (OMP) \cite{OMP01} are classified into the second category, which needs partial priori information of the channel, e.g., the degree of sparsity $L$. ITD-SE based Least Square Estimation (LSE) needs the fewer iterations, but its performance depends on the adaptive threshold selection scheme. ACS leverages the advanced compressed sensing theory and combines with the hybrid beamforming technique. Therefore, it is very suitable for the mmWave systems. The algorithms in the third category do not need any priori knowledge of the channel distribution except the noise variance, e.g., Sparse Bayesian Learning (SBL) \cite{SBL}, LASSO \cite{LASSO}, Expectation-maximization Bernoulli-Gaussian Approximate Message Passing (EM-BG-AMP) \cite{P_Schniter_EM}, etc. However, most of these algorithms need to learn the channel in order to improve the estimation performance, for example, SBL that is implemented via the more robust T-MSBL \cite{T_SBL}. In addition, although LASSO \cite{LASSO} and EM-BG-AMP \cite{P_Schniter_EM} have the lower complexity, their solutions are generally not the sparsest, and EM-BG-AMP also requires independent and identically distributed (i.i.d.) zero-mean Gaussian training matrix.  \cite{P_Schniter_EM,J_Mo2014}. Recently,  \cite{Guo_Qing} proposed a modified mean field (MF) message passing-based algorithm, which also belongs to the third category and can deliver even better performance with the lower complexity than the conventional vector-form MF SBL algorithm by introducing a few hard constraint factors. This also provides a promising method for the future mmWave channel estimation.


In this paper, we develop an iterative channel estimation algorithm based on the LSE, Expectation-Maximization (EM) and Sparse Message Passing (SMP) for mmWave MIMO systems with large antenna arrays at both the transmitter and receiver. A beamspace channel representation model is adopted which can capture the sparseness of physical mmWave channel and provides simple geometric interpretation of the scatter environment ([see \cite{P_Schniter_virtual,Sayeed_SP2002}]). Based on this representation, we further model the mmWave channel as a Bernoulli-Gaussian distribution. We summarize our main contributions of this paper as follows.
\begin{itemize}
  \item We formulate a sparse channel estimation problem and propose a novel sparse channel estimation algorithm. Compared with existing sparse channel estimation methods, ours can yield a better performance since it not only can take full advantage of the inherent sparseness of the mmWave channel, but also can leverage both virtues of the LSE and SMP algorithms.
  \item We give the performance analysis of the proposed algorithm, and derive its upper bound and CRLB. Furthermore, we show that the proposed algorithm is the Minimum Variance Unbiased Estimator (MVUE) under the assumption that we have the partial priori knowledge of the channel.
  \item We employ the Extrinsic Information Transfer (EXIT) chart-based technique for the convergence analysis of the key part of the proposed algorithm, and provide insights on the iteration evolution of the proposed algorithm, based on which design parameters we optimize.  This analysis use a Gaussian approximation for message densities under the density evolution, and adopt the Log-Likelihood Ratios (LLRs) of messages, which can reduce the complexity of the analysis. 
  \item We evaluate the performance of the proposed estimation algorithm. Numerical simulations show that our algorithm exhibits far better performance than the classical LSE estimator, as well as existing sparse channel estimators (e.g., LASSO, ITD-SE, EM-BG-AMP, etc.). In addition, we also find that this algorithm can approximately achieve the CRLB with the fast convergence speed.
\end{itemize}

The rest of the paper is organized as follows. In Section II, we present the beamspace channel representation model for mmWave MIMO systems and formulate a sparse channel estimation problem. In Section III, we propose an iterative sparse channel estimation algorithm. The performance analysis of the proposed algorithm is given in Section IV.  In Section V, simulation results demonstrating the performance of the proposed algorithms are given, before concluding the paper in Section VI.

\textit{Notation}: $ a $ is a scalar, $ \mathbf{a} $ is a vector and $ \mathbf{A} $  is a matrix. $ \mathbf{A}^T$, $ \mathbf{A}^H$, $ \mathbf{A}^{-1}$, $ \mathbf{A^\dag}$ and $ \|\mathbf{A}\|_F $  represent transpose, Hermitian (conjugate transpose), inverse, pseudo-inverse and Frobenius norm of a matrix $ \mathbf{A} $, respectively. $ \mathbf{A} \otimes \mathbf{B} $ denotes the Kronecker product of $ \mathbf{A} $ and $ \mathbf{B} $, and $ vec(\mathbf{A}) $  is a vector stacking all the columns of $ \mathbf{A} $. $ diag(\mathbf{a}) $ is a diagonal matrix with the entries of $ \mathbf{a} $ on its diagonal, and $ diag(\mathbf{A}) $ is a block diagonal matrix with the matrix $\mathbf{A}$ as the block on its diagonal. $\mathcal{N}(\mathbf{x};\mathbf{m},\mathbf{V}) $ is the Probability Distribution Function (PDF) of a complex Gaussian random vector $\mathbf{x}$ with mean $\mathbf{m}$ and covariance $ \mathbf{V}$. We use the $E\{\cdot \}$, $ Var\{\cdot \}$, $ \mathrm{exp}(\cdot )$, and $\mathbb{R}$ to denote the expectation, variance, nature exponential operation  and the real field respectively. In addition, $E\{a|b\}$ denotes the conditional expectation of variable $a$  given $b$, and  $Var\{a|b\}$ denotes the conditional variance of the variable $a$ given $b$.

\begin{figure*}
  \begin{center}
  \includegraphics[width=152mm]{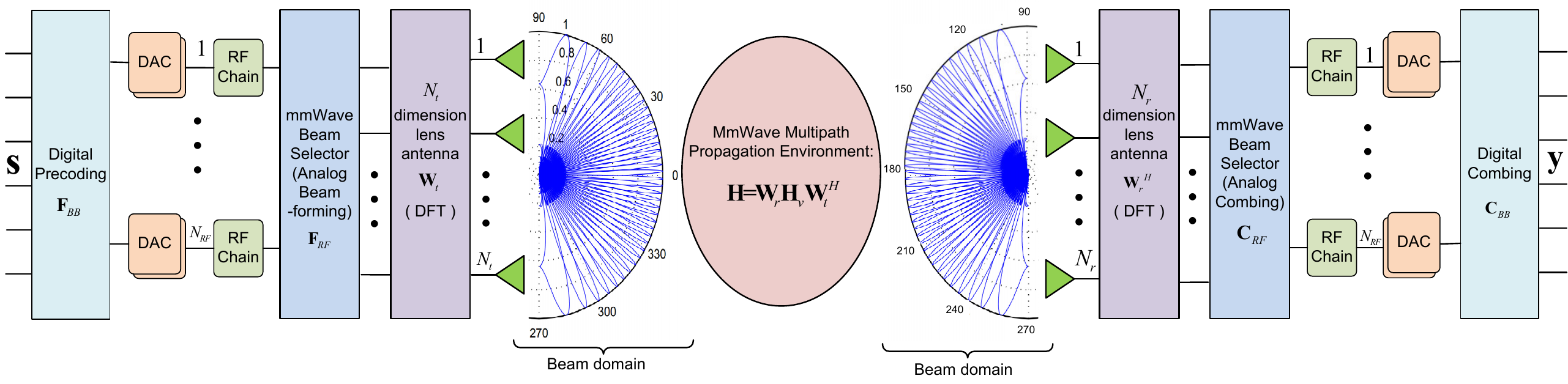}  \vspace{-4mm}%
  \caption{\!\!\! The hybrid analog-digital communication architecture based on the beamspace channel representation.  Essentially, the beamspace channel representation is to map the signal of the spatial domain to the signal of the beam domain by employing a carefully designed discrete lens antenna array instead of the electromagnetic antenna array. The Fourier transformation $ \mathbf{W}_{t}$ and $ \mathbf{W}_{r}$  can be seen as a mapping from the antenna domain onto a beam domain and the entries of the matrix $ \mathbf{H}_v$ can be interpreted as the channel gains between the $ N_t$ transmit and the $ N_r $ receive beams.}
  \label{fig:Virtual_Channel} \vspace{-8mm}
  \end{center}
\end{figure*} 
\section{SYSTEM MODEL}
We consider a hybrid analog-digital mmWave communication system that has $ N_t $ transmit  and $ N_r $  receive antennas at the transmitter and receiver respectively, and both of them have $ N_{RF} $ RF chains \footnote{For simiplicity, we assume the same number of RF chains at the transmitter and receiver. The proposed architecture also can be extended to the case where has different numbers of RF chains.}. 
The transmitter and receiver communicate via $N_s$ data streams, such that $ N_s \leq N_{RF} \leq N_{r} $ and $N_s \leq N_{RF} \leq N_{t}$ \cite{Alk_channel_Estimation,J_Mo2014}. Assuming frequency-flat fading channel, and that there is a $N_{RF} \times N_s$ baseband precoder $\mathbf{F_{BB}}$ followed by an $N_t \times N_{RF}$ RF precoder  $\mathbf{F_{RF}}$ in the downlink transmission, we can denote $\mathbf{F}= \mathbf{F_{RF}F_{BB}}$  as a $N_t \times N_s$ combined precoding matrix, and similarly, we denote $\mathbf{C}$ as the $N_r \times N_s$ combining matrix, which is composed of the RF combiners $\mathbf{C_{RF}}$ and baseband combiners $\mathbf{C_{BB}}$. For the traditional hybrid analog-digital model, the observed signal at the receiver can be written as \cite{Alk_channel_Estimation,J_Mo2014,D_Tse_books,sparse_MIMO_Channel,Alkhateeb_limited_feedback}, \vspace{-1mm}

\begin{equation}\label{model_1}
  \mathbf{y}=\mathbf{C^{H}HFs}+\mathbf{C^{H}z},
\end{equation}
where $ \mathbf{H }\in \mathbb{C}^{N_r \times N_t}$ is the channel matrix, $ \mathbf{s} \in \mathbb{C}^{N_s \times 1}$ is the transmitted signal, $ \mathbf{y }\in \mathbb{C}^{N_s \times 1}$ is the received signal£¬ and $ \mathbf{z} \in \mathbb{C}^{N_r \times 1} $ is the Gaussian noise with $\mathbf{z} \sim \mathcal{N}(0, \sigma^2_{n}\mathbf{I})$ .

Since mmWave channels are expected to have limited scattering, we adopt a geometric channel model with $L$ scatterers. Each scatterer is further
assumed to contribute a single propagation path between transmitters and receivers \cite{rappaport_mmWave_model2015,Alk_channel_Estimation,J_Mo2014}. Under this model, the channel  $ \mathbf{H } $ can be expressed as
\begin{equation}\label{2}
\mathbf{ H}=\sqrt{\frac{N_{r}N_{t}} {\rho}} \sum_{l=1}^L\alpha_l\mathbf{a}_{r}(\theta_l)\mathbf{a}_{t}^H(\phi_l),
\end{equation}
where $\rho$ denotes the average path-loss between the transmitter and receiver, $ \alpha_l$ is the gain of the $ l $th path, $ \phi_l\in[0,2\pi]$ and $ \theta_l\in[0,2\pi]$ denote the $l$th path's azimuth angles of departure and arrival of the transmitter and receiver respectively. Finally,  $ \mathbf{a}_{t}(\phi_l)$ and $ \mathbf{a}_{r}(\theta_l)$  are the antenna array response vectors at the transmitter and receiver respectively \cite{Alk_channel_Estimation,D_Tse_books,sparse_MIMO_Channel,Alkhateeb_limited_feedback}. If a uniform linear arrays is used, $ \mathbf{a}_{t}(\phi_l) $ can be written as
\begin{equation}\label{3}
\mathbf{\mathbf{a}}_{t}(\phi_l)\!=\!\frac{1}{\sqrt{N_{t}}}\!\!\left[1,e^{j\frac{2\pi}{\lambda}dsin(\phi_l)},...,e^{j(N_{t}-1)\frac{2\pi}{\lambda}dsin(\phi_l)}\right]^T,
\end{equation}
where $ \lambda$ is the signal wavelength, and $ d $ is the distance between antenna elements. The array response vectors at the receiver, $ \mathbf{a}_{r}(\theta_l)$, can be written in a similar fashion. Then, the channel can be written in a more compact form as
\begin{equation}\label{4}
  \mathbf{H}=\mathbf{A}_{r}diag( \bm{\alpha})\mathbf{A}_{t}^H,
\end{equation}
where $ \bm{\alpha} =\sqrt{\frac{N_{r}N_{t}} {\rho}}[\alpha_1,\alpha_2,...,\alpha_l]^T.$ The matrices \\ \vspace{-2mm}
\begin{equation}\label{5}
  \mathbf{A}_{t}=[\mathbf{a}_{t}(\phi_1),\mathbf{a}_{t}(\phi_2),...,\mathbf{a}_{t}(\phi_l)],
\end{equation}
and \vspace{-0mm}
\begin{equation}\label{6}
\mathbf{A}_{r}=[\mathbf{a}_{r}(\theta_1),\mathbf{a}_{r}(\theta_2),...,\mathbf{a}_{r}(\theta_l)],
\end{equation} contain the transmitter and receiver array response vectors.  
The mmWave multipath propagation channel usually consists of a few reflected path clusters \cite{A.Ghosh_LOS,J_Mo2014,Alkhateeb_limited_feedback,multipath_clustering}.  Large antenna arrays are deployed in the mmWave systems for combatting the high path loss. Hence, we usually have $ L \ll min \{N_{r},N_{t}\}$.

For capturing the inherent sparse characteristic of the physical mmWave modeling, we adopt a hybrid analog-digital communication architecture that is based on the beamspace channel representation as shown in Fig. \ref{fig:Virtual_Channel}. This beamspace representation also provides a tractable linear channel characterization, and offers a simple and transparent interpretation to the effects of scattering and array characteristics on channel capacity and diversity \cite{Sayeed_SP2002,Sayeed_beamspace,Sayeed_Differential_beamspace,Dai_linglong_01,Sayeed_beamspace_02}. Essentially, the beamspace channel representation is to map the signal of the spatial domain to the signal of the beam domain by employing a carefully designed discrete lens antenna array instead of the electromagnetic antenna array \cite{veug03,Sayeed_beamspace,Sayeed_Differential_beamspace,Dai_linglong_01,Sayeed_beamspace_02,beamspace_01}. In particular, the lens acts as a virtual passive phase shifter, focusing the incident electromagnetic wave to a certain region. This lens, when used jointly with antennas, exhibits two significant properties: (i) focused signal power at the front end achieving high directivity and gain, and (ii) concentrated signal power directed to a sub-region of the antenna array. These properties
make the lens a practical and energy efficiency tool for implementing the RF frontend in beamforming systems \cite{hcw_ICASSP_2018,zengyong_01,zengyong_02,hcw_globecom_2018}. The finite dimensionality of the signal space allows the beamspace channel model that can be expressed as
\begin{equation}\label{7}
\begin{split}
\mathbf{H}_v =\mathbf{W}_{r}^H\mathbf{H}\mathbf{W}_{t}
\end{split}
\end{equation}
where  $ \mathbf{W}_{r} \in \mathbb{C}^{N_r \times N_r }$ and $\mathbf{W}_{t} \in \mathbb{C}^{N_t \times N_t }$ are channel-invariant unitary DFT matrices \cite{Sayeed_beamspace,P_Schniter_virtual_sparse}, and $ \mathbf{W}_{t}\mathbf{W}_{t}^H=\mathbf{I}_{N_t}, \mathbf{W}_{r}\mathbf{W}_{r}^H=\mathbf{I}_{N_r} $.
Note that  $\mathbf{H}_v \in \mathbb{C}^{N_r\times N_t}$ is  no longer diagonal. We recast (\ref{model_1}) by the beamspace channel representation as \vspace{-2mm}
\begin{equation}\label{8}
\begin{split}
\mathbf{y}=\mathbf{C^{H}}\mathbf{W}_{r}\mathbf{H}_v\mathbf{W}_{t}^{H}\mathbf{F}\mathbf{s}+\mathbf{C^{H}}\mathbf{W}_{r}\mathbf{z},
\end{split}
\end{equation}

The Fourier transformation $ \mathbf{W}_{t}$ and $ \mathbf{W}_{r}$  can be seen as a mapping from the antenna domain onto a beam domain and the entries of the matrix $ \mathbf{H}_v$ can be interpreted as the channel gains between the $ N_t$ transmit and the $ N_r $ receive beams\cite{ Sayeed_beamspace}. Assuming the channel is time-invariant in the blocks ${t \in \{1,...,T \} }$. Then, $\mathbf{Y}\triangleq[\mathbf{y}_{1},...,\mathbf{y}_{T}] $, $\mathbf{S}\triangleq[\mathbf{\mathbf{s}}_1,...,\mathbf{s}_T] $
, and $\mathbf{N}\triangleq[\mathbf{C^{H}}\mathbf{\mathbf{W}_{r}\mathbf{z}}_{1},...,\mathbf{C^{H}}\mathbf{W}_{r}\mathbf{z}_{T}]\triangleq [\mathbf{n}_1,...,\mathbf{n}_T] $. The channel model is rewritten as
\begin{equation}\label{9}
\begin{split}
\mathbf{Y}=\mathbf{C^{H}}\mathbf{W}_{r}\mathbf{H}_v\mathbf{W}_{t}^{H}\mathbf{F}\mathbf{S}+\mathbf{N},
\end{split}
\end{equation}
where $ \mathbf{Y}\in \mathbb{C}^{N_s \times T} $, $ \mathbf{S} \in \mathbb{C}^{N_s \times T} $, and $ \mathbf{N} \in \mathbb{C}^{N_s \times T} $. We define  $\mathbf{D}\triangleq \mathbf{C^{H}}\mathbf{W}_{r}$ and $ \mathbf{X}\triangleq \mathbf{W}_{t}^{H}\mathbf{F} \mathbf{S}$. Then, we recast the (\ref{9}) as
\begin{equation}\label{10}
\begin{split}
\mathbf{Y} =\mathbf{D}\mathbf{H}_v\mathbf{X}+\bm{\mathbf{N}}.
\end{split}
\end{equation}
Then,  vectorizing (10) \cite{Alk_channel_Estimation} yields
\begin{equation}\label{11}
\begin{split}
vec(\mathbf{Y})&=vec(\mathbf{D}\mathbf{H}_v\mathbf{X}) + vec(\mathbf{N})\\
& =( \mathbf{X}^T \otimes \mathbf{D}   )vec(\mathbf{H}_v)+vec( \mathbf{N}).
\end{split}
\end{equation}  
By defining the $\mathbf{\bar{y}}\triangleq vec(\mathbf{Y})$, $ \mathbf{\bar{S}}\triangleq \mathbf{X}^T \otimes \mathbf{D} $, $\mathbf{h}_v \triangleq vec(\mathbf{H}_v)$ and $\mathbf{\bar{n}} \triangleq vec( \mathbf{N})$, (\ref{11}) is equivalently rewritten as follows: \vspace{-1mm}
\begin{equation}\label{12}
\begin{split}
\mathbf{\bar{y}}=\mathbf{\bar{S}}\mathbf{h}_v+\mathbf{\bar{n}},
\end{split}
\end{equation}
where  $ \mathbf{\bar{y}} \in \mathbb{C}^{N_sT \times 1} $, $ \mathbf{\bar{S}} \in \mathbb{C}^{N_sT \times N_rN_t} $, $ \mathbf{h}_v \in \mathbb{C}^{N_rN_t \times 1} $ and $ \mathbf{\bar{n}} \in \mathbb{C}^{N_sT \times 1} $. 
The mmWave channel estimation problem is simplified to estimate the beamspace channel vector $ \mathbf{h}_v $ by the equivalent training matrix $ \mathbf{\bar{S}} $ and the observed vector $ \mathbf{\bar{y}} $.

\section{SPARSE CHANNEL ESTIMATION}
In this section, we  present an iterative channel estimation algorithm based on the LSE and SMP algorithm as shown in Fig. 2, which is named LSE-SMP. It consists of four phases: LSE Coarse Estimation (Step 1), Sparse Message Passing Detection (Step 2), Update for LSE Estimation and Sparsity Ratios (Step 3), and Decision and Output (Step 4). Since there is no priori knowledge of $ \mathbf{h}_v$, we initially adopt the LSE method to obtain its coarse estimation. Then, in the Step 2, we consider the estimation of non-zero positions in the channel vector $ \mathbf{h}_v $ as a detection problem, and propose an SMP algorithm to find these non-zero positions under the estimated sparsity ratio. In the Step 3, we apply LSE method again by leveraging the estimated non-zero positions in the Step 2 to obtain the fine estimation of $ \mathbf{h}_v$, and estimate the sparsity ratio by the Expectation-Maximization (EM) method. The fourth Step is to make the decision according to the  performance requirements (MSE, number of iterations, etc.) and the estimation of the Step 2 and Step 3. If the MSE of LSE-SMP meets the requirement or the number of iterations reaches the limit, the final estimation of $ \mathbf{h}_v $ will be output, otherwise the Step 2 and Step 3 will repeat until we obtain a fine estimation.

\subsection{LSE Coarse Estimation}
To find the Coarse Estimation $ \mathbf{\hat{h}}_v $ based on the observed vector $\mathbf{\bar{y}}$ with a Mean Square Error (MSE) $ E\{\| \mathbf{\hat{h}}_{v}-\mathbf{h}_{v}\|^{2}\} $, we can solve the following Least Square (LS) problem,
\begin{equation}\label{13}
\begin{split}
\mathbf{\hat{h}}_v &=\arg \min_{\mathbf{h}_v}\|\mathbf{\bar{y}}-\mathbf{\bar{S}}\mathbf{h}_v\|^{2}_2 \\
&=[\mathbf{\bar{S}}^H\mathbf{\bar{S}}]^{-1}\mathbf{\bar{S}}^H\mathbf{\bar{y}}.
\end{split}
\end{equation}\vspace{0.5mm}

It is noted that it is MVUE in the sense of MSE for the deterministic signal when the estimator does not have any prior knowledge about either the sparsity structure of $ \mathbf{h}_v$ (i.e., the distribution and location of non-zero entries), or its degree of sparsity (i.e., $L$).
\begin{figure}
  \begin{center}
  \includegraphics[width=83mm]{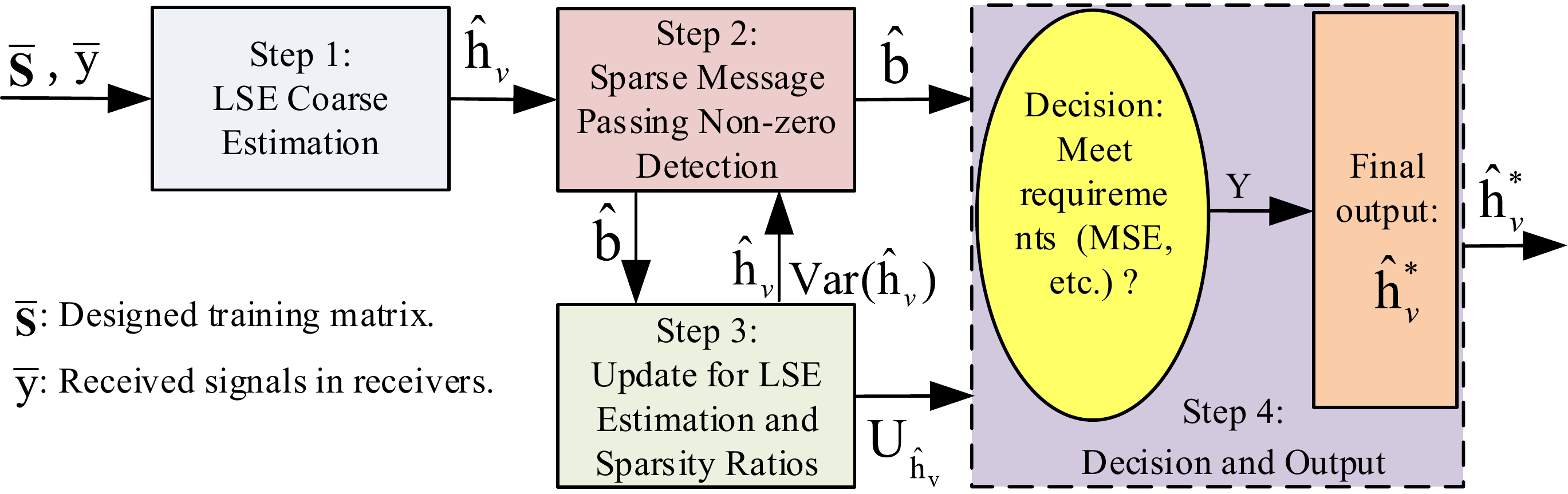}  \vspace{-2mm} %
   \caption{The processing for the proposed LSE-SMP algorithm, which consists of four phases: LSE Coarse Estimation, Sparse Message Passing Detection, Update for LSE Estimation and Sparsity Ratios, and Decision and Output.}
  \label{fig:2} \vspace{-6mm}
  \end{center}
\end{figure}
\subsection{Sparse Message Passing Algorithm}
After we get the Coarse Estimation of $ \mathbf{h}_v $, we propose a fast iterative algorithm to find the positions of non-zero entries. This algorithm is named sparse message passing since it can take full advantage of the channel sparsity and message passing algorithm.
\subsubsection{Factor Graph Representation of the mmWave Channel}
In order to get better understanding of our proposed algorithm, we show the factor graph representation of the channel vector $ \mathbf{h}_v $ in the following. Firstly, we decompose the $ \mathbf{h}_v $ into a diagonal coefficient matrix $ \mathbf{U}_{\mathbf{h}_{v}} $ and a column array $ \textbf{b} $. The column array $ \textbf{b}=[b_{ij} ]_{N_rN_t \times 1} ( i \in \{0,...,N_r\},  j \in \{0,...,N_t\} ) $ is called the position vector, and it represents the positions of non-zero in the coefficient matrix $ \mathbf{U}_{\mathbf{h}_{v}} $. The $ b_{ij}\in\{1,0\}$ can be seen as a Bernoulli distribution. Then, the $ \mathbf{h}_v $ can be recast as
\begin{figure*}
  \begin{center}
  \includegraphics[width=141mm]{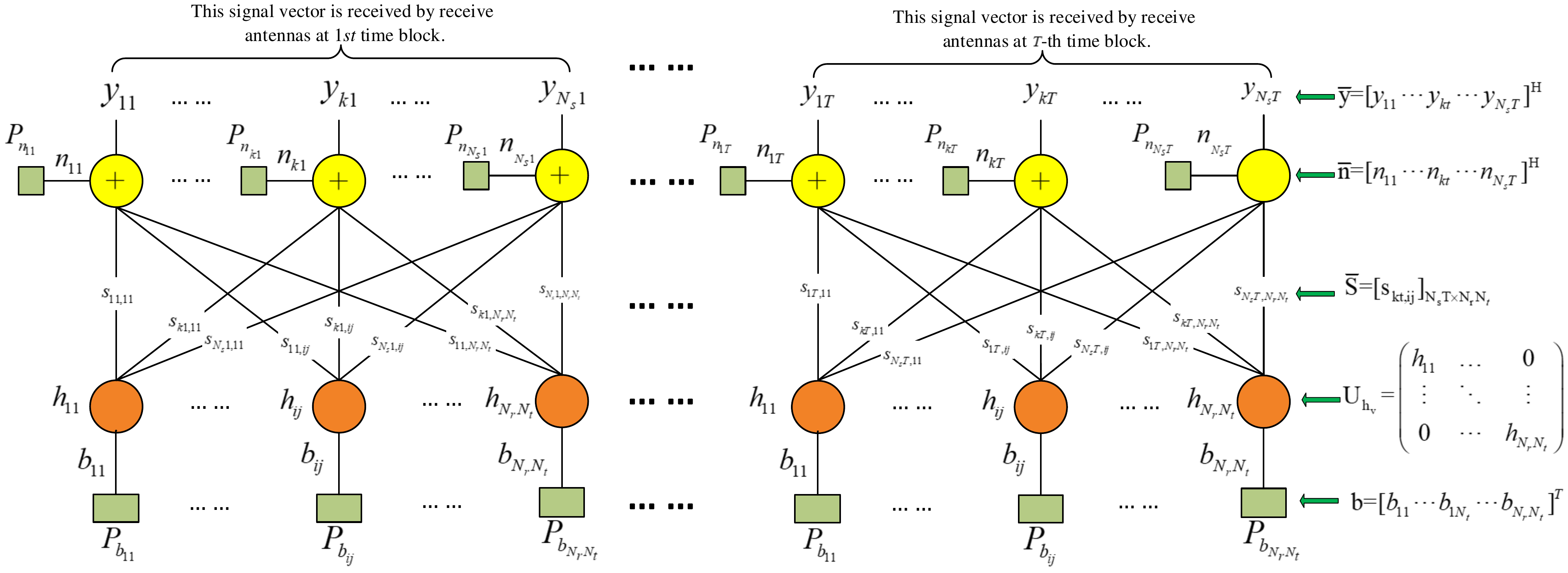} \vspace{-3mm}  \\%
  \caption{The factor graph representation for the proposed sparse message passing detection algorithm. This factor graph is plotted on the basis of equations (\ref{14})-(\ref{18}) and factor graph rules.  The nodes ($ n_{11},...,n_{N_rT}$) and ($ h_{11},...,h_{N_rN_t}$) are named the sum and variable nodes respectively.}
  \label{fig:All_Factor_Graph} \vspace{-8mm}
  \end{center}
\end{figure*} \vspace{-1mm}
\begin{equation}\label{14}
\begin{split}
\mathbf{h}_v
 = \underbrace{ \left[\begin{array}{ccccc}
    h_{11}  &           &          &      &   0          \\
            &   \ddots  &          &                 \\
            &           & h_{1N_t} &                \\
            &           &          &    \ddots   &     \\
     0      &           &          &              &   h_{N_rN_t}      \\
\end{array}\right]}_{= \mathbf{U}_{\mathbf{h}_v}}
\underbrace{\left[\begin{array}{cccc}
  b_{11} \\
  \vdots \\
  b_{1N_t} \\
  \vdots \\
  b_{N_rN_t}
\end{array}\right]}_{=\mathbf{b}}
\end{split}.
\end{equation}

Then, we rewrite (\ref{12}) as
\begin{equation}\label{17}
\begin{split}
\!\!\!\!\!\!\underbrace{\left[\begin{array}{ccccc}
  y_{11} &  \ldots  & y_{1T}  &  \ldots  &  y_{N_sT}
\end{array}\right]^H}_\mathbf{\bar{y}} &= \mathbf{\bar{S}}\mathbf{h}_v +\mathbf{\bar{n}} \\
                   &=\mathbf{\bar{S}} \mathbf{U}_{\mathbf{h}_v} \mathbf{b}+\mathbf{\bar{n}}.
\end{split}
\end{equation}

According to factor graph analysis rules \cite{Loeliger_factor_graph,factor_graph_LDPC,P_Schniter_facror_graph,Lei2015_TWC}, we can plot the factor graph to represent above equations, and it is shown in the Fig. \ref{fig:All_Factor_Graph}. The nodes ($ n_{11},...,n_{N_rT}$) and ($ h_{11},...,h_{N_rN_t}$) are named the sum and variable nodes respectively.

The proposed SMP algorithm is considered for estimating positions of non-zero entries. It is similar to the belief propagation decoding process of the low density parity check code, in which the output message called extrinsic information on each edge is calculated by the messages on the other edges that are connected with the same node \cite{factor_graph_LDPC,Li_ping_factor_graph,CHEMP2014,Lei2015,LDPC_Guassion}.

\subsubsection{Message Update at Sum Nodes }
\begin{figure}
  \begin{center}
  \includegraphics[width=84 mm]{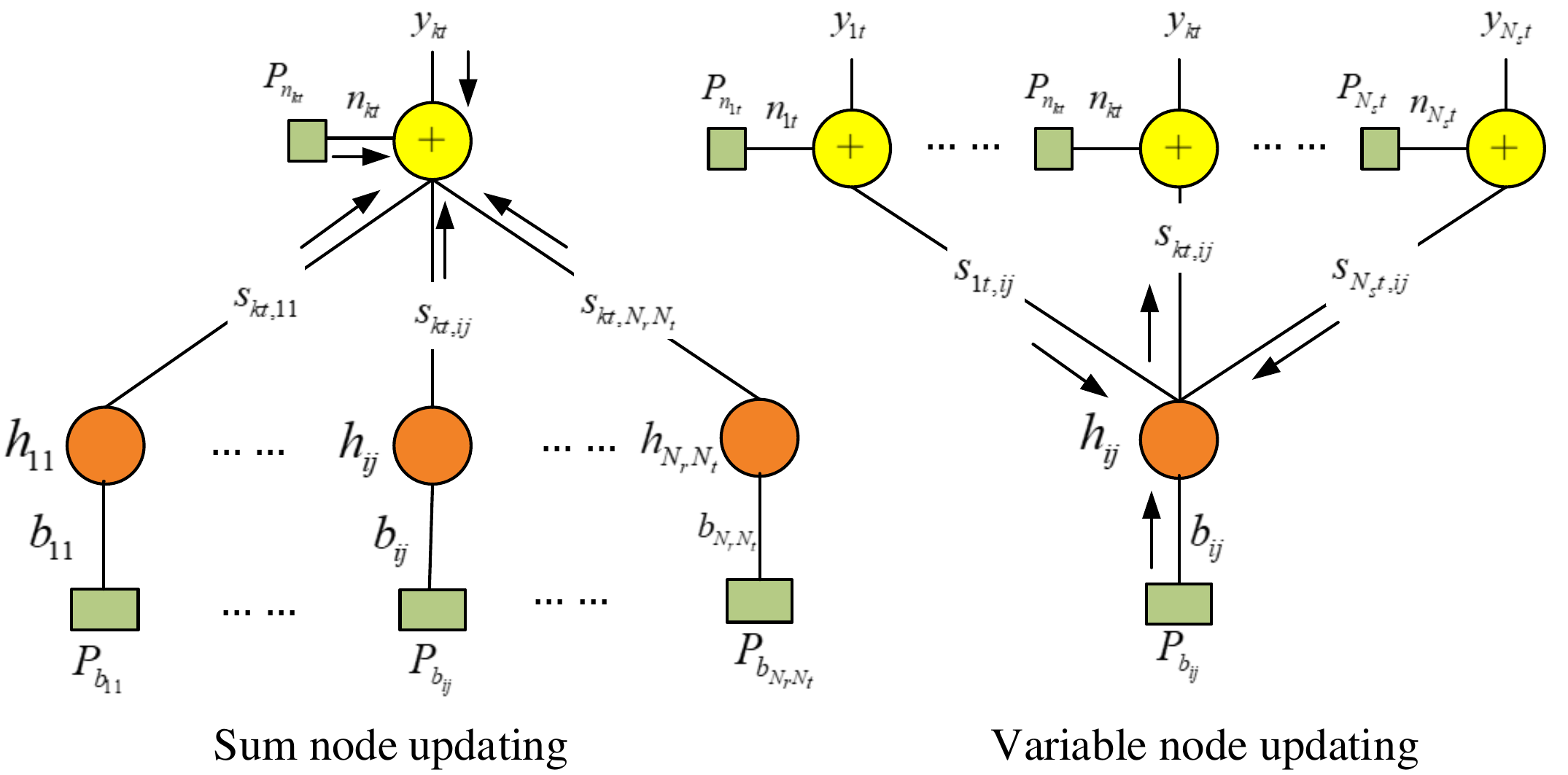} \vspace{-2mm}\\  %
  \caption{Messages update at sum nodes and variable nodes. The output message called extrinsic information on each edge is calculated by the messages on the other edges that are connected with the same node. For the Gaussian-Bernoulli sparse signals, the messages passing on each edge are the probabilities of a Bernoulli distribution. The mean and variance of a Gaussian distribution are updated at the sum nodes, and they are used for computing the probability of a Bernoulli distribution. }
  \label{fig:Sum_Variable_node_undating} \vspace{-6mm}
  \end{center}
\end{figure}

To analyze a sum node that is shown in the Fig. \ref{fig:Sum_Variable_node_undating}, we can obtain the $k$th ($ k \in \{0,...,N_s\}$) data stream at the $ t $th ($ t \in \{0,...,T\}$) time block, and it can be expressed as \cite{LDPC_Richard,hcwllei_globecom_2016,hcw_TVT_2018} \vspace{-0mm}
\begin{equation}\label{18}
\begin{split}
y_{kt}= \sum_{i=1}^{N_r}\sum_{j=1}^{N_t}s_{kt,ij}h_{ij}b_{ij}+n_{kt}.
\end{split}
\end{equation} \vspace{-0mm}
As we mentioned before, there are  $ L $ non-zero entries in the vector of $ \textbf{b} $. Then, we have the definition of the sparsity ratio $\eta = \frac{L}{N_rN_t}$. Assuming that the entries of $\mathbf{b}=[b_{11},...,b_{N_tN_r}]$ are i.i.d., we can know the probability of the Bernoulli distribution, which can be denoted as
\begin{numcases}{}
p_0(b_{ij}=1)= \hat{\eta}, \label{19} \\
p_0(b_{ij}=0)=1-\hat{\eta}, \label{19_1}
\end{numcases}                
where $\hat{\eta}$ is the estimation of $\eta$, and it will be updated by the EM method in the next phase. It should be pointed out that the initial value of $\eta$ is set as 0.5. When $ N_t $ or $ N_r $ goes very large, the term $ \sum_{i=1}^{N_r}\sum_{j=1}^{N_t}s_{kt,ij}h_{ij}b_{ij} $ can be approximated as the Gaussian distribution \cite{Lei2015_TWC,LDPC_Guassion,analysis_for_MP} according to the law of large numbers. When we compute the probability of $ p(b_{ij}=1) $ from  the $ kt $ sum node to the $ ij $ variable node, we consider the messages from the other variable nodes $ \ell m $ ($ \ell \neq i , m\neq j $ and $\ell \in \{1,2,...,N_r\}, m\in \{1,2,...,N_t\}$) to the sum node $ kt $ as the \textbf{equivalent Gaussian noise} $n^{\ast}_{kt} $. This can be expressed as \vspace{-0mm}
\begin{equation}\label{20}
\begin{split}
\!\!\!\!y_{kt}\!\!= \!\!\underbrace{s_{kt,ij}h_{ij}p(b_{ij}\!=\!1)}_{\mathrm{Desired \,\,Item}} \!+\underbrace{\sum_{\ell \neq i}^{N_r}\!\sum_{m\neq j}^{N_t}s_{kt,\ell m}h_{\ell m}p(b_{\ell m}\!=\!1)+\!n_{kt}}_{{\mathrm{Equivalent\, Gaussian \,noise}: \,n^{\ast}_{kt}}}.
\end{split}
\end{equation}

Furthermore, we can compute the mean value and variance of the equivalent Gaussian noise $n^{\ast}_{kt} $\cite{hcwllei_globecom_2017,analysis_for_MP}. These messages update at the sum nodes are given  by
\begin{equation}\label{21}
\begin{split}
\vspace{-1cm}\!\!\!\!\!\!\!\! \!\!\!\!\!\!\!\!\!\!\!\!\!\!\!\!\!\!\!\!\!\!\!\!\!\!\!\!\!\!\!\!\!\!\!\!\!\!\!\!\!\!\!\!\!\!\!\!\!\!\!\!\!\!\!\!e^{s}_{kt\rightarrow ij}(\tau) &=E \left\{ n^{\ast}_{kt}|\mathbf{s}_{kt},\hat{\mathbf{h}}(\tau),\mathbf{p}^{v}(\tau) \right\} \\
&=\sum\limits_{\ell \neq i}\sum\limits_{m\neq j}s_{kt,\ell m}\hat{h}_{\ell m}(\tau)p^{v}_{\ell m \rightarrow kt}(\tau),
\end{split}
\end{equation}
\begin{equation}\label{21_1}
\begin{aligned}
v^{s}_{kt\rightarrow ij}(\tau)&= Var\left\{n^{\ast}_{kt}|\mathbf{s}_{kt},\hat{\mathbf{h}}(\tau),\mathbf{p}^{v}(\tau),\hat{\mathbf{v}}_{h}(\tau)\right\}\\ 
&=\sum\limits_{\ell \neq i}\sum\limits_{m\neq j}s^{2}_{kt,\ell m}E\left\{\hat{h^2}_{\ell m}(\tau)\right\}E\left\{{p^{v}}^2_{\ell m \rightarrow kt}(\tau)\right\}\\
&\quad\,\, -s^2_{kt,\ell m}\hat{h^2}_{\ell m}(\tau){p^{v}}^2_{\ell m \rightarrow kt}(\tau) \\
&=\sum\limits_{\ell \neq i}\!\sum\limits_{m\neq j}\!s^{2}_{kt,\ell m}p^{v}_{\ell m \rightarrow kt}(\tau)\hat{h}_{\ell m}^{2}(\tau)(1\!-\!p^{v}_{\ell m \rightarrow kt}(\tau)) \\
&\quad\,\, +s^{2}_{kt,\ell m}p^{v}_{\ell m \rightarrow kt}(\tau)v_{h_{\ell m}}(\tau) +\sigma^{2}_{n}.
\end{aligned}
\end{equation}
where $\tau$ denotes the iteration number, and $\sigma^{2}_{n}$ is the variance of the Gaussian noise, and $\mathbf{s}_{kt}=[s_{kt,11},...s_{kt,ij},...,s_{kt,N_rN_t}]^H $, $\hat{\mathbf{h}}=[\hat{h}_{11},...\hat{h}_{ij},...,\hat{h}_{N_rN_t}]^H $, $\mathbf{p}^{v}=[p^{v}_{11 \rightarrow kt},..., p^{v}_{ij \rightarrow kt},...,p^{v}_{N_rN_t \rightarrow kt}]^T $ and $\hat{\mathbf{v}}_{h}=[v_{h_{11}},...,[v_{h_{ij}},..., v_{h_{N_rN_t}}]^H $.
In addtion, $ e^{s}_{kt\rightarrow ij}(\tau) $ and $ v^{s}_{kt\rightarrow ij}(\tau)$ denote the mean and variance of the equivalent Gaussian noise $n^{\ast}_{kt} $ when the message $ s_{kt,ij}h_{ij}p(b_{ij}=1)$ passes from the $ kt $ sum node to the $ ij$ variable node at the $\tau $th iteration. Similarly, $ p^{v}_{\ell m \rightarrow kt}(\tau) $ denotes the probability message of $ p(b_{\ell m}=1) $ passing from the $ \ell m $ variable node to the $  kt $ sum node at the $ \tau $th iteration. $ \hat{h}_{\ell m} $ and $ v_{h_{\ell m}} $ that denote the mean and variance of  $ h_{\ell m} $ are estimated in the Step 3.
Once we obtain the mean and variance of the equivalent Gaussian noise, we can compute the statistical probability of $b_{ij}=1$ and $b_{ij}=0$ as below equations in accordance to $ \left( y_{kt}-s_{kt,ij}h_{ij}p(b_{ij}=1) \right) \sim \mathcal{N }(e^{s}_{kt\rightarrow ij}(\tau),v^{s}_{kt\rightarrow ij}(\tau))$.
\begin{numcases}{}   
P(b_{ij}=1|y_{kt},s_{kt,ij}(\tau),\hat{h}_{ij}(\tau),v_{h_{ij}}(\tau))  \nonumber\\
\,=\mathcal{N}(y_{kt};e^{s}_{kt\rightarrow ij}(\tau)+s_{kt,ij}\hat{h}_{ij}(\tau),v^{s}_{kt\rightarrow ij}(\tau)+s^{2}_{t j}v_{h_{ij}}(\tau)), \nonumber\\
P(b_{ij}=0|y_{kt},s_{kt,ij}(\tau),\hat{h}_{ij}(\tau),v_{h_{ij}}(\tau)) \nonumber\\
\,=\mathcal{N}(y_{kt};e^{s}_{kt\rightarrow ij}(\tau),v^{s}_{kt\rightarrow ij}(\tau)). \label{22}
\end{numcases}

Then, we can give the probability message of $b_{ij}=1$ passing from the $ kt $ sum node to the $  ij $ variable node as follows
\begin{equation}   
\begin{split}
&p^{s}_{kt\rightarrow ij}(\tau)\!=\!\left(\!1+\!\frac{P(b_{ij}=0|y_{kt},s_{kt,ij}(\tau),\hat{h}_{ij}(\tau),v_{h_{ij}}(\tau))}{P(b_{ij}=1|y_{kt},s_{kt,ij}(\tau),\hat{h}_{ij}(\tau),v_{h_{ij}}(\tau))}\right)^{-1} \\
&=\Bigg(1+\sqrt{\frac{v^{s}_{kt\rightarrow ij}(\tau)+s^{2}_{t j}v_{h_{ij}}(\tau)}{v^{s}_{kt\rightarrow ij}(\tau)}}\mathrm{exp}\Bigg( \frac{(y_{kt}-e^{s}_{kt\rightarrow ij}(\tau))^2}{-2v^{s}_{kt\rightarrow ij}(\tau)} \\
&+\frac{(y_{kt}-e^{s}_{kt\rightarrow ij}(\tau)-s_{kt,ij}\hat{h}_{ij}(\tau))^2}{2(v^{s}_{kt\rightarrow ij}(\tau)+s^{2}_{t j}v_{h_{ij}}(\tau))}\Bigg)\Bigg)^{-1}. \label{23}
\end{split}
\end{equation}
\subsubsection{Message Update at Variable Nodes }
In terms of the message update at variable nodes, we consider variable nodes as a broadcast process \cite{Li_Ping_maltiple_access,Lei2016_TVT} and the message update at the variable node is given by
\begin{equation}\label{24}    
\begin{split}
\!\!p^{v}_{ij \rightarrow kt}(\tau \!+\!1)\!= \!\Bigg(\!1+\!\frac{\prod\limits_{\kappa \neq k}p^{s}_{\kappa t \rightarrow ij}(\tau)\cdot p_0(b_{ij}=0)}{\prod\limits_{\kappa \neq k}p^{s}_{\kappa t \rightarrow ij}(\tau)\cdot p_0(b_{ij}=1)}\Bigg)^{-1},
\end{split}
\end{equation}
where $\kappa \in \{1,...,N_s\} $, and $ p^{s}_{kt \rightarrow ij}(\tau +1) $ denotes the probability message of $b_{ij}=1$ passing from the $ it $ sum node to the $  ij $ variable node at the $ (\tau +1)$th iteration. Furthermore, we can obtain the estimation of the Bernoulli variable $ b_{ij}$ at the $ (\tau +1)$th iteration as
\begin{equation}\label{28} \vspace{-0mm}
\begin{split}
\hat{b}_{ij}(\tau +1)=\Bigg(1+ \frac{\prod\limits_{k=1}^{N_s}p^{s}_{kt \rightarrow ij}(\tau)\cdot p_0(b_{ij}=0)}{\prod\limits_{k=1}^{N_s}p^{s}_{kt \rightarrow ij}(\tau)\cdot p_0(b_{ij}=1)}\Bigg)^{-1}.
\end{split}
\end{equation}

\textbf{\textit{Remark 1:}} It should be pointed out that $ p^{s}_{kt \rightarrow ij}(\tau +1) $ is the extrinsic information and will be used to update messages of the sum nodes in the next iteration. On the other hand, $\hat{b}_{ij}(\tau +1)$ is updated based on the full information coming from all the sum nodes, and it will be used in the Step 3 for the estimation of $\mathbf{h}_v$.

\textbf{\textit{Remark 2:}} The proposed SMP that is based on the message passing can obtain near optimal performance without heavy computational complexity by approximating the term $ \sum_{i=1}^{N_r}\sum_{j=1}^{N_t}s_{kt,ij}h_{ij}b_{ij} $ as the Gaussian distribution. This is because the number of transmit antennas $N_t$ is large in mmWave communication systems as they operate at the higher communication spectrum, therefore, the approximation is accurate due to the law of large numbers. This also shows that our proposed algorithm is specialized for mmWave systems.

\subsection{Update for LSE Estimation and Sparsity Ratio} \vspace{-0.0mm}

\subsubsection{LSE Fine Estimation}
Once the positions of the non-zero entries have been estimated,  the next step is to estimate the value of the coefficient matrix $\mathbf{U}_{\mathbf{h}_v} $. For the problem, we propose a novel strategy based on the LSE method. This strategy is to \textbf{ swap the positions } of $h_{ij}$ and $ b_{ij}$ in the (\ref{14}), so that we can get an accurate estimation by leveraging the sparsity of $\textbf{b}$. Rewriting  (\ref{17}) as
\begin{equation}\label{25}   
\mathbf{\bar{y}} =\mathbf{\bar{S}} \mathbf{U_{\hat{b}}} \mathbf{h}_v+\mathbf{\bar{n}},
\end{equation}
where $\mathbf{h}_v \in \mathbb{C}^{N_tN_r \times 1 }$, and  $ \mathbf{\hat{b}}=[\hat{b}_{ij} ]_{N_rN_t \times 1}$ is the vector estimated  by the SMP algorithm. Similar with the LSE Coarse Estimation, the estimation of $\mathbf{h}_v$ can be obtained by solving the following LS problem
\begin{equation}\label{26}   
\mathbf{\hat{h}}_v = \arg \min_{\mathbf{h}_v}\{ \| \mathbf{\bar{y}}-\mathbf{\bar{S}} \mathbf{U_{\hat{b}}} \mathbf{h}_v\|^{2}_2\}.
\end{equation}
Solving the above expression, we get the following estimator for $ \mathbf{h}_v$ as
\begin{numcases}{}
\mathbf{\hat{h}}_v(\tau)
=\mathbf{\hat{Q}}^{\dagger}(\tau)\left(\mathbf{\bar{S}}\mathbf{U}_{\mathbf{\hat{b}}}(\tau)\right)^H\mathbf{\bar{y}}, \label{27} \\
\mathbf{\hat{v}}_h(\tau)
=\sigma^{2}_n\left(\left(\mathbf{\bar{S}}\mathbf{U}_{\mathbf{\hat{b}}}(\tau)\right)^H \mathbf{\bar{S}}\mathbf{U}_{\mathbf{\hat{b}}}(\tau)\right)^{\dagger}, \label{27_1}
\end{numcases}
where $\mathbf{\hat{Q}}(\tau)=\mathbf{U}_{\mathbf{\hat{b}}}(\tau)\mathbf{\bar{S}}^H\mathbf{\bar{S}}\mathbf{U}_{\mathbf{\hat{b}}}(\tau) $, $ \mathbf{\hat{v}}_h=\mathrm{diag } [v_{h_{i j}}]_{N_rN_t \times N_rN_t}$. $ \mathbf{\hat{h}}_v(\tau) $ and $\mathbf{\hat{v}}_h(\tau)$  denotes the estimated value and variance of $ \mathbf{h}_v$ at $\tau$th iteration. After we obtain $ \mathbf{\hat{h}}_v(\tau) $ and $\mathbf{\hat{v}}_h(\tau)$, these values will replace the $ \mathbf{\hat{h}}_v(k-1) $ and $\mathbf{\hat{v}}_h(k-1)$  for calculating the mean and variance of the equivalent Gaussian noise $n^{\ast}_{kt} $ in the iteration. \vspace{-0mm}

\subsubsection{EM update for the sparsity ratio }
We now present a classical EM algorithm \cite{P_Schniter_EM,EM_alogrithm} to learn the sparsity ratio $\eta $.  Since the channel vector $\mathbf{h}_v$ can be modeled as i.i.d Bernoulli-Gaussian, then we have the marginal PDF as
\begin{equation}\label{18_0} \vspace{1mm}
p(h_{ij};\eta,u_{h_v},v_{h_v} )=(1-\eta)\delta(h_{ij})+\eta \mathcal{N}(h_{ij};u_{h_v},v_{h_v}),
\end{equation}
\hspace{-0.1mm}where $\delta(\cdot)$ is the Dirac delta, and $u_{h_v}$ and $v_{h_v}$ are the mean and variance of non-zero entries of $h_{ij}$ respectively. Then, we can give the EM update for $\eta$ by the estimated the parameters $\mathbf{\hat{r}}(\tau)\triangleq [\hat{\eta}(\tau),\hat{u}_{h_v}(\tau),\hat{v}_{h_v}(\tau)]$. In the sequel, we use the $ \mathbf{\hat{r}}_{\backslash \eta}$ to denote the vector $\mathbf{\hat{r}}$ with the element $\eta$ removed. Similar to the equation (29) in paper \cite{P_Schniter_EM}, the EM update for $\eta$ can be written as
\begin{equation}\label{18_1} \vspace{-1mm}
\hat{\eta}(\tau +1) \!=\!\arg \!\max_{\eta \in (0,1)} \!\sum_{j=1}^{N_t}\sum_{i=1}^{N_r} E\{ \mathrm{ln}\,\,p(h_{ij};\eta,\mathbf{\hat{r}}_{\backslash \eta}(\tau)|\bar{\mathbf{y}}; \mathbf{\hat{r}}(\tau))\}.
\end{equation} \vspace{-0mm}
To maximize the value of $\eta$ in the above equation, it is necessary to zero the derivative of the sum; i.e., that satisfies
\begin{equation}\label{18_2} \vspace{-0mm}
\sum_{j=1}^{N_t}\sum_{i=1}^{N_r} \int_{h_{ij}} p(h_{ij}|\bar{\mathbf{y}}; \mathbf{\hat{r}}(\tau))\frac{d\,\,\mathrm{ln}\,\,p(h_{ij};\eta,\mathbf{\hat{r}}_{\backslash \eta}(\tau))}{d \eta}=0.
\end{equation} \vspace{-0mm}
Finally, we get the solution of (\ref{18_2}) that is
\begin{equation}\label{18_3} \vspace{-0mm}
\hat{\eta}(\tau +1)\!=\!\frac{1}{N_tN_r}\sum_{j=1}^{N_t}\sum_{i=1}^{N_r} \frac{1}{1\!+\!\left(\frac{\eta(\tau)\mathcal{N}(r;\ \hat{h}_{ij}(\tau),\hat{v}_{h_{ij}}(\tau)+\mu^r)}{(1-\eta(\tau))\mathcal{N}(r;\,0,\mu^r)}\right)^{-1}},
\end{equation} \vspace{-0mm}
\hspace{-0.2mm}where $r=\frac{1}{N_s}\sum_{k=1}^{N_s}s_{kt,ij}^{-1}(y_{kt}-e^{s}_{kt\rightarrow ij}(\tau))$, $\mu^r=\frac{1}{N_s}\sum_{k=1}^{N_s}s_{kt,ij}^{-2}v^{s}_{kt\rightarrow ij}(\tau)$, and $r$ can be interpreted as a $\mu^r$-variance-AWGN corrupted observation of the true $h_{ij}$. Their detailed derivations are similar with the equation (9) and (15) of the paper \cite{P_Schniter_EM}.
\subsection{ Decision and Output of LSE-SMP }
When the MSE of the LSE-SMP meets the requirement or the number of iterations reaches the limit, we output the final estimation of channel vector $\mathbf{h}_{v}$ as
\begin{numcases}{}
\mathbf{\hat{h}}_v=\mathbf{\hat{Q}}^{\dagger}(\tau) \left(\mathbf{\bar{S}}\mathbf{U}_{\mathbf{\hat{b}}}(\tau)\right)^H\mathbf{\bar{y}}, \label{29}\\
\mathbf{\hat{h}}_{v}^*=\mathbf{U}_{\mathbf{\hat{h}}_v}\mathbf{\hat{b}}, \label{29_1}
\end{numcases}
where $\mathbf{\hat{b}}= [\hat{b}_{ij}]_{N_rN_t \times 1} $. It should be pointed out that the final output is based on the SMP and LSE Fine Estimation.

\textbf{\textit{Remark 3:}} The LSE fine estimation makes full use of the sparse information that is estimated by the SMP at each iteration, which not only can significantly improve estimation performance of LSE, but also accelerate the whole algorithm to converge. In turn, the estimated result of LSE is used to improve the estimation accuracy of SMP in the next iteration. In other words, the LSE and SMP will help each other at each iteration for improving the performance of estimation until the MSE approaches the minimum or meets the system requirement. When the channel $\mathbf{h}_v$ is more sparse, the advantage of the proposed algorithm becomes significant. Therefore, this is the another reason that the proposed algorithm is suitable for mmWave communication systems. 

\subsection{ LLRs of LSE-SMP}
From (\ref{24}) and (\ref{28}), we notice that the messages update for the variable nodes are easy to overflow in the simulations due to the multiplications of a large number of probabilities. Therefore, we use Log-Likelihood Ratios (LLRs) scheme \cite{LDPC_Guassion,Sten_Exit_Chart} to replace the computation of the non-zero probabilities during the message update process. This not only can prevent the overflow, but also can reduce the complexity of computation. The LLRs scheme can be written as follows
\begin{numcases}{}
l^{s}_{ kt \rightarrow ij }(\tau)=\mathrm{log}\frac{p^{s}_{  kt \rightarrow ij }(\tau)}{1-p^{s}_{  kt \rightarrow ij }(\tau)}, \label{30} \\
l^{v}_{ ij \rightarrow kt}(\tau)=\mathrm{log}\frac{p^{v}_{  ij \rightarrow kt }(\tau)}{1-p^{v}_{ ij \rightarrow kt }(\tau)},\label{30_1} \\
l_{0}=\mathrm{log}\frac{P_0(b_{ ij}=1)}{1-P_0(b_{ ij}=1)}, \label{30_2}
\end{numcases}
for any $ i \in \{1,...,N_r\}$, $ j\in \{1,...,N_t\}$ , $
k \in \{1,...,N_s\}$ and $\tau$ ($\tau$ denotes the number of iterations ). $ l^{s}_{kt \rightarrow ij}(\tau)$ denotes the LLRs of the probability message of $b_{ij}=1$ passing from the $ kt $ sum node to $  ij $ variable node. Similarly, $ l^{v}_{ ij \rightarrow kt}(\tau)$ denotes the LLRs of the probability message of $b_{ij}=1$ passing from the $ ij $ variable node to $ kt $ sum node. Then, the LSE-SMP algorithm is updated by LLRs as follows.
\subsubsection{ Message Update at Sum Nodes }
Based on (\ref{21}), (\ref{21_1}) and (\ref{23}), the LLRs of Bernoulli-Gaussian messages updating at sum nodes are rewritten as

\begin{equation}\label{31_2}
\begin{split}
&l^{s}_{kt\rightarrow ij}(\tau)=-\mathrm{log} \left(\sqrt{\frac{v^{s}_{kt\rightarrow ij}(\tau)+s^{2}_{t j}v_{h_{ij}}(\tau)}{v^{s}_{kt\rightarrow ij}(\tau)}}\right)  \\
&-\frac{(y_{kt}-e^{s}_{kt\rightarrow ij}(\tau)-s_{kt,ij}\hat{h}_{ij}(\tau))^2}{2(v^{s}_{kt\rightarrow ij}(\tau)+s^{2}_{t j}v_{h_{ij}}(\tau))} + \frac{(y_{kt}-e^{s}_{kt\rightarrow ij}(\tau))^2}{2v^{s}_{kt\rightarrow ij}(\tau)}.
\end{split}
\end{equation}\label{35}
\subsubsection{ Message Update at Variable Nodes }
Based on (\ref{24}) and (\ref{28}),  the LLRs of messages updating at variable nodes are rewritten as \vspace{-0.0cm}
\begin{numcases}{}
l^{v}_{ij \rightarrow kt}(\tau +1)=l_{0}+\sum_{\kappa \neq k}^{N_s}l^{s}_{kt\rightarrow ij}(\tau), \label{32} \\
l^{v}_{ij}(\tau +1)=l_{0}+\sum \limits_{k=1}^{N_s}l^{s}_{kt\rightarrow ij}(\tau), \label{33_1} \\
\hat{b}_{ij}(\tau +1)=  1/(1+e^{-l^{v}_{ij}(\tau +1)}). \label{33}
\end{numcases}
where the message $l^{v}_{ij}(\tau +1)$ is updated based on the full information coming from all the sum nodes. It is used for calculating the $\hat{b}_{ij}(\tau +1)$. 

From (\ref{31_2})-(\ref{33}), it should be noticed that the LSE-SMP algorithm in the LLRs form will be more concise, which can reduce the complexity of computation and prevent the overflow of multiplications of a large number of probabilities, since LLRs transform the multiplication operations into the addition operations. In addition, we also can see that this will be convenient to analyze and predict the performance of the system in the following section IV.

\subsection{ LSE-SMP in the Matrix Form }

In order to reduce the complexity of matrix inversions and multiplications, we can split the high dimension diagonal matrix and block diagonal matrix into some low dimension matrixs, since the transmit and receive antennas are independent each other as we mentioned before. As we defined before, we denote $ i\in \{1,2,...,N_r\}$, $j\in \{1,2,...,N_t\} \, $,$  \,k \in \{1,2,...,N_s\} $and  $ \,t \in \{1,2,...,T\} $, and $k$ denotes the $\tau$th iteration in the following definitions. Then, we give some definitions as follows: $\mathbf{U}^s(\tau)=\left[e_{kt \rightarrow ij}^s(\tau)\right]_{N_sT \times N_rN_t} $, $ \mathbf{V}^s(\tau)=\left[v_{kt \rightarrow ij}^s(\tau)\right]_{N_sT \times N_rN_t} $, $\mathbf{P}^s(\tau)=\left[p_{kt \rightarrow ij}^s(\tau)\right]_{N_sT \times N_rN_t} $,
$ \mathbf{L}^s(\tau)=\left[l_{kt \rightarrow ij}^s(\tau)\right]_{N_sT \times N_rN_t} $,
$ \mathbf{P}^v(\tau)=\left[p_{ij \rightarrow kt }^v(\tau)\right]_{N_rN_t \times N_sT }$, $\mathbf{L}^v(\tau)=\left[l_{ij \rightarrow kt }^v(\tau)\right]_{N_rN_t\times N_sT }$,
$\mathbf{U}_{\hat{\mathbf{h}}_v}(\tau)= \mathrm{diag}[\hat{h}_{ij} (\tau)]_{N_r N_t \times 1 }$, $ \mathbf{U}_{\mathbf{\hat{b}}}(\tau)=\mathrm{diag}\left[\hat{b}_{ij}(\tau) \right]^T_{N_r N_t \times 1} $, $ \mathbf{\hat{V}}_{h}(\tau)= \mathrm{diag}\left[v_{h_{ij}}(\tau) \right]_{N_r N_t \times 1} $, $\mathbf{\bar{y}}= \left[y_{kt} \right]^H_{N_sT \times 1 } $, $\mathbf{\hat{Q}}(\tau)=\mathbf{U}_{\mathbf{\hat{b}}}(\tau)\mathbf{\bar{S}}\mathbf{\bar{S}}^H\mathbf{U}_{\mathbf{\hat{b}}}(\tau) $.

In addition, we let $\mathbf{A}_{N_r \times N_t}.*\mathbf{B}_{N_r \times N_t}=[a_{ij}b_{ij}]_{N_r \times N_t}$, $\mathbf{A}_{N_r \times N_t}^{(2)}=[a_{ij}^{2}]_{N_r \times N_t}$, $\mathbf{1}_{N_r \times N_t}=[1]_{N_r \times N_t}$, and $\mathbf{C}_{N_r \times N_r}=\mathbf{A}_{N_r \times N_t}\cdot \mathbf{B}_{N_t \times N_r} $ is the matrix product of $\mathbf{A}$ and $\mathbf{B}$.

Then, the algorithm 1 shows the detailed process of the LSE-SMP in the LLRs matrix form.
\begin{algorithm}
\caption{ LSE-SMP Algorithm}
\begin{algorithmic}[1]
\vspace{-0.0cm}
\State {\small{\textbf{Input:} {{$\mathbf{S}$, $ T, N_s, N_r, N_t, N_{kte} $, $\sigma^2_n$, $\epsilon>0$, $\tau=0$,}} initial value $ \eta=0.5$, and calculate {{$\mathbf{\bar{S}}^{(2)}$}} and  $(\mathbf{\bar{S}}^H\mathbf{\bar{S}})^{-1}$,
\vspace{-0.0cm}
\State \textbf{Initialized Coarse LSE Estimation:} 
\vspace{-0.0cm}
\, $\mathbf{\hat{V}}_{h}(0)= \sigma_n^2[\mathbf{\bar{S}}\mathbf{\bar{S}}^H]^{-1} $, $\mathbf{P}^v(0)= \mathbf{0}$, $\mathbf{U}_{\mathbf{\hat{h}}}(0)$ and $\mathbf{L}^v(0)= \mathbf{0}$.
\vspace{-0.00cm}
\State \textbf{Do}
\vspace{-0.0cm}
\vspace{-0.0cm}
\State { \quad\,$\widetilde{\mathbf{V}}^*(\tau) =  \Big( \mathbf{U}_{\mathbf{\hat{h}}}^{(2)}(\tau) \cdot \mathbf{P}^v(\tau).* \left(\mathbf{1}_{ N_sT \times  N_rN_t  }-\mathbf{P}^v(\tau) \right) $ \\
\quad\,\, $ + \mathbf{\hat{V}}_{h}(\tau) \cdot \mathbf{P}^v(\tau)\Big).*\mathbf{\bar{S}}^{(2)H}$, $\widetilde{\mathbf{U}}^*(\tau) = \mathbf{U}_{\mathbf{\hat{h}}_v}(\tau)\cdot \mathbf{\bar{S}}^H .*\mathbf{P}^v(\tau)  $,\\
\quad\,\,\, and $\mathbf{P}^v(\tau)=\big(\mathbf{1}_{    N_rN_t  \times N_sT }+e^{-\mathbf{L}^v(\tau)}\big)^{-1}$,
\vspace{-0.0cm}
\State  \vspace{-0.22cm}\[
\begin{array}{l}
\quad\,\,\left[ \!\!\begin{array}{l}
\mathbf{U}^s(\tau)\\
\mathbf{V}^s(\tau)
\end{array} \!\!\right] \!=\! \left[ \!\!\begin{array}{c}
 \widetilde{\mathbf{U}}^{*}(\tau) \cdot \mathbf{1}_{N_sT \times 1}  \mathop {}\limits_{\mathop { }}\\
\sigma_n^2\cdot\mathbf{1}_{N_rN_t \times 1}\!+\! \widetilde{\mathbf{V}}^{*}(\tau) \cdot \mathbf{1}_{N_sT \times 1}
\end{array}\!\!\! \right]\! \cdot\! {\mathbf{1}_{1 \times N_sT}} \\
\qquad\qquad\quad\quad\quad\,\,-\left[ \!\!\begin{array}{c}
\widetilde{\mathbf{U}}^{*^H}(\tau) \mathop {}\limits_{\mathop { }}\\
 \widetilde{\mathbf{V}}^{*^H}(\tau)
\end{array}\!\!\! \right],
\end{array}\]}}
\State\vspace{-0.22cm}\[ \!\!\!\!\begin{array}{c}
\qquad\!\!\! \begin{array}{c}
\mathbf{L}^s(\tau)
\end{array} \!\!= \!\!\begin{array}{c}
{\frac{1}{2}}\log\frac{\mathbf{V}^s(\tau)+\mathbf{\bar{S}}^{(2)}\cdot\mathbf{\hat{V}}_{h}(\tau)}{\mathbf{V}^s(\tau)}+\frac{\left(\mathbf{\bar{y}} \cdot {\mathbf{1}_{1 \times N_rN_t}} -\mathbf{U}^s(\tau)\right)^2}{2\mathbf{V}^s(\tau)} \end{array} \!\!\! \\
\,\,\,\,\,\,\,\,\, \!\!\!\begin{array}{c}
-\frac{\left(\mathbf{\bar{y}} \cdot {\mathbf{1}_{1 \times N_r N_t}}-\mathbf{U}^s(\tau)-  \mathbf{\bar{S}} \cdot \mathbf{U}_{\mathbf{\hat{h}}_v}(\tau) \right)^2}{2\left(\mathbf{V}^s(\tau)+\mathbf{\bar{S}}^{(2)}\cdot\mathbf{\hat{V}}_{h}(\tau)\right)}
\end{array} \!\!\!\!,
\end{array} \]
\State\vspace{-0.22cm}\[ \quad \begin{array}{c}
\left[\!\!\! \begin{array}{c}
\mathbf{L}^v(\!\tau \!+\!1\!) \\
\mathbf{L}(\!\tau \!+\!1\!) \\
\mathbf{\hat{b}}(\!\tau \!+\!1\!)
\end{array} \!\!\!\right]\!\! = \!\!\left[ \!\!\!\begin{array}{c}
 \mathbf{1}_{N_sT \times 1} \!\cdot \!\left[ \mathbf{1}_{1\times N_sT}\cdot \mathbf{L}^s(\tau) \right]\!-\!\mathbf{L}^s(\tau) \mathop {}\limits_{\mathop { }} \\
\quad \mathbf{1}_{N_sT \times 1} \cdot \left[ \mathbf{1}_{1\times N_sT}\cdot \mathbf{L}^s(\tau) \right] \mathop {}\limits_{\mathop { }} \\
\quad \mathbf{1}_{N_rN_t \times N_s T}/(\mathbf{1}_{N_rN_t \times N_s T}+e^{-\mathbf{L}(\!\tau \!+\!1\!)}) \\
\end{array} \!\!\!\right] \\
\hspace{-1.9cm}+ \left[ \!\!\!\begin{array}{c}
 \left(\mathbf{L}^s(0)\right)^{H}\\
\left(\mathbf{L}^s(0)\right)^{H} \mathop {}\limits_{\mathop { }} \\
  0
\end{array} \!\!\!\right].
\end{array}\]
\State \vspace{-0.22cm}\[\!\!\!\!\!\!\!\!\!\!\!\!\!\!\!\!\!\!\!\!\!\!\!\! \begin{array}{c}
\hspace{-0.0cm}\left[\!\!\! \begin{array}{c}
\mathbf{\hat{h}}_v(\tau) \\
\mathbf{\hat{V}}_{h}(\tau)
\end{array} \!\!\!\right]\!\! =\!\!\left[ \!\!\! \!\!\!\begin{array}{c}
\quad  \mathbf{\hat{Q}}{^{\dagger}(\tau)} \left(\mathbf{\bar{S}}\mathbf{U}_{\mathbf{\hat{b}}(\tau)}\right)^H\mathbf{\bar{y}}  \mathop {}\limits_{\mathop { }}\\
\quad \quad \sigma^{2}_n\left(\mathbf{\hat{Q}}{^{\dagger}(\tau)}\right)^{\dagger}
\end{array} \!\!\!\right]\!\!,
\end{array}\]
\vspace{-0.32cm}
\vspace{0.00cm}
\State \hspace{0 cm} \quad $\hat{\eta}(\tau +1)\!=\!\frac{1}{N_tN_r}\sum\limits_{j=1}^{N_t}\sum\limits_{i=1}^{N_r} \frac{1}{1\!+\!\left(\frac{\eta(\tau)\mathcal{N}(r;\ \hat{h}_{ij}(\tau),\hat{v}_{h_{ij}}(\tau)+\mu^r)}{(1-\eta(\tau))\mathcal{N}(r;\,0,\mu^r)}\right)^{-1}}$,
\State \quad\,\,     $\tau = \tau +1$,
\vspace{0.12cm}
\State  \textbf{While} \;{\small{$\big( \:(\|\mathbf{U}_{\mathbf{\hat{h}}_v}(\tau +1)-\mathbf{U}_{\mathbf{\hat{h}}_v}(\tau)\|_2<\epsilon \,\,\& \,\, \|\mathbf{L}(\tau +1)-\mathbf{L}{(\tau)}\|_2<\epsilon)
\;{\textbf{or}}\; \tau \leq N_{kte} \;\big)$}}
\vspace{-0.0cm}
\State  \vspace{-0.2cm}{{ \[\!\!\!\!\begin{array}{c}
\vspace{-0.0cm}
\quad \,\,\left[\!\!\! \begin{array}{c}
\mathbf{\hat{h}}_v \\
\mathbf{\hat{h}}_{v}^*
\end{array} \!\!\!\right]\!\! \!=\left[ \!\!\! \!\!\!\begin{array}{c}
\quad \mathbf{\hat{Q}}^{\dagger}(\mathbf{\bar{S}}\mathbf{U}_{\mathbf{\hat{b}}})^H \mathbf{\bar{y}}  \mathop {}\limits_{\mathop { }}\\
\quad \quad \mathbf{U}_{\mathbf{\hat{h}}_v}\mathbf{\hat{b}}
\end{array} \!\!\!\right]\!\!,
\end{array}\qquad\qquad\qquad\qquad\qquad\qquad\qquad\qquad\qquad\qquad\;\]}}
\State \textbf{Output:} $\mathbf{\hat{h}}_v$  and  $\mathbf{\hat{h}}_{v}^*$.}
\vspace{-0.12cm}
\end{algorithmic}
\end{algorithm} \vspace{-0cm}

\section{ PERFORMANCE ANALYSIS }

\subsection{ Cram\'{e}r-Rao Low Bound Of LSE-SMP }
In this section, we give the analysis of CRLB and show that our proposed LSE-SMP algorithm is unbiased under under the assumption that we have the partial priori knowledge of the channel. Firstly, we consider the case that the channel vector $\mathbf{h}_v$ is a deterministic and non-sparse. From  \cite{CRLB_book,CRLB_book2} and the signal model in (\ref{13}), we can yield the CRLB of the conventional LSE as
\begin{equation}\label{40}
\mathbf{CRLB_{LSE}} \fp \geq \mathbf{C_{LSE}}=\sigma^{2}_n(\mathbf{\bar{S}}^H \mathbf{\bar{S}})^{-1},
\end{equation}
where $\mathbf{C_{LSE}}$ is the covariance matrix of $\mathbf{h}_v$ for the LSE estimation. Note that LSE is the MVUE, and the detailed proof can be found in \cite{CRLB_book,CRLB_book2}. 
Compared with the non-sparse case, the sparse case is slightly more complex. In particular, we are interested in the lower bound for the estimation accuracy, however, the proposed LSE-SMP algorithm involves a big loop that contains the LSE estimation, solving Gaussian functions, SMP estimation and EM learning, hence it is extremely difficult to establish an analytical model. Recalling the in (\ref{14}), we know that $\mathbf{h}_v$ can be decomposed into the two parts ($\mathbf{U}_{\mathbf{h}_v}$ and $\mathbf{b}$). Therefore, we can take the way like alternating minimization method \cite{Alternating_Minimization01,R_Niazadeh_sparse,Alternating_Minimization02} to analyze these two parts independently. In other words, when we analyze the estimated performance of $\mathbf{U}_{\mathbf{h}_v}$, we assume that the $\mathbf{b}$ is given, and vice versa. Therefore, we have the following assumption 1.

\vspace{-0cm}
\textbf{Assumption 1}: The channel $\mathbf{h}_v$ is deterministic and we have the priori knowledge of non-zero positions $\mathbf{b}$.

The paper \cite{Lei2015_TWC,MSE_Chart} had already shown that the massage passing-based algorithm can converge to the LMMSE under the Gaussian distribution from the theoretical perspective. In addition, we also perform a few simulations under the typical settings, and simulations show that the estimation for non-zero positions is accurate, especially with small number of non-zero entries. These simulation results was attached as the supplemental materials. Under the Assumption 1, we have

\textbf{Theorem 1:} The proposed LSE-SMP is the MVUE, and can achieve the CRLB that is given by
\begin{equation}\label{41}   
\begin{split}
\mathbf{CRLB_{LSE-SMP}} \fp \geq \sigma^{2}_n \left(\left(\mathbf{\bar{S}}\mathbf{U}_{\mathbf{b}}\right)^H\mathbf{\bar{S}}\mathbf{U}_{\mathbf{b}}\right)^{\dag}.
\end{split}
\end{equation}
\begin{IEEEproof}
The first step is to verify that the our proposed estimator is unbiased  under the Assumption
1. Recalling the signal model in (\ref{25}) and the definition of the unbiased estimator, we get
\begin{align}\label{42}
\!\!& E(\mathbf{\hat{h}}_v) \!=\!\!E\left\{\left(\left(\mathbf{\bar{S}}\mathbf{U}_{\mathbf{b}}\right)^H\mathbf{\bar{S}}\mathbf{U}_{\mathbf{b}}\right)^\dag \left(\mathbf{\bar{S}}\mathbf{U}_{\mathbf{b}}\right)^H \mathbf{\bar{y}}\right\} \nonumber \\
& \!\!=\!\!E\left\{\left(\left(\mathbf{\bar{S}}\mathbf{U}_{\mathbf{b}}\right)^H\mathbf{\bar{S}}\mathbf{U}_{\mathbf{b}}\right)^\dag \left(\mathbf{\bar{S}}\mathbf{U}_{\mathbf{b}}\right)^H (\mathbf{\bar{S}}\mathbf{U}_{\mathbf{b}} \mathbf{h}_v+\mathbf{\bar{n}})\right\} \nonumber \\
&\!\!=\!\!\mathbf{U}_{\mathbf{b}}\mathbf{h}_v=\mathbf{h}_v.
\end{align}  \vspace{-0.0cm}
So, it is a unbiased estimator. The next step is to compute its CRLB and verify that our proposed LSE-SMP algorithm can achieve the CRLB under the Assumption
1. As previously mentioned, the additive white Gaussian noise is modeled as $ \mathcal{N}(0,\sigma^{2}_n\mathbf{I})$ and the channel  $ \mathbf{h}_v $ is a deterministic vector.  Recalling the signal model in (\ref{12}), we can get
\begin{equation}\label{43}   
\mathbf{\bar{y}} \sim p(\mathbf{\bar{y}}|\mathbf{h}_v)=\mathcal{N}(\mathbf{\bar{S}}\mathbf{h}_v,\sigma^{2}_n\mathbf{I}),
\end{equation}
where $p(\mathbf{\bar{y}}|\mathbf{h}_v)$ is the probability density function of $\mathbf{\bar{y}}$ under the condition of known $\mathbf{h}_v$.
Then, we can compute the $ \frac{ \partial ln \,p(\mathbf{\bar{y}}|\mathbf{h}_v)}{\partial \mathbf{h}_v } $,
\begin{equation}\label{44}   
\frac{ \partial ln \,p(\mathbf{\bar{y}}|\mathbf{h}_v)}{\partial \mathbf{h}_v }
   \!\! =\!\!\frac{1}{\sigma_n^{2}}\left[\left(\mathbf{\bar{S}}\mathbf{U}_{\mathbf{b}}\right)^H\mathbf{\bar{y}}-\left(\mathbf{\bar{S}}\mathbf{U}_{\mathbf{b}}\right)^H \mathbf{\bar{S}}\mathbf{U}_{\mathbf{b}}\mathbf{h}_v \right].
\end{equation}
Then, we obtain the following expression for the Fisher Information Matrix (FIM):
\begin{equation}\label{45}   
I(\mathbf{h}_v)=-E\left\{\frac{ \partial^2 ln\, p(\mathbf{\bar{y}}|\mathbf{h}_v)}{\partial \mathbf{h}^{2}_v}\right\} \\
           = \frac{1}{\sigma_n^{2}}{{\left(\mathbf{\mathbf{\bar{S}}}\mathbf{U}_{\mathbf{b}}\right)^H\mathbf{\bar{S}}}\mathbf{U}_{\mathbf{b}}}.
\end{equation}

We note that $ \mathbf{\bar{S}} \mathbf{U}_{\mathbf{b}}$ has the rank no larger than $L$  due to the multiplication of $ \mathbf{\bar{S}} $ by $ \mathbf{U}_{\mathbf{b}}$. The matrices $ \mathbf{\bar{S}} \mathbf{U}_{\mathbf{b}} $ and $\left(\mathbf{\bar{S}} \mathbf{U}_{\mathbf{b}}\right)^H $ have some all zero columns (and rows), so it is singular. For this type singular matrix, it need to meet a constraint \cite{CRLB_kay}, otherwise our proposed estimator (\ref{29_1}) has infinite variance that renders the CRLB useless. Before we analyze this constraint, we firstly compute the following key identity
\begin{equation}\label{46}   
\begin{split}
\mathbf{G}
&=\left(\left(\mathbf{\bar{S}}\mathbf{U}_{\mathbf{b}}\right)^H\mathbf{\bar{S}}\mathbf{U}_{\mathbf{b}}\right)^\dag\left(\mathbf{\bar{S}}\mathbf{U}_{\mathbf{b}}\right)^H\mathbf{\bar{S}}\mathbf{U}_{\mathbf{b}} \\
&= diag(\mathbf{b})\neq \mathbf{I}_{N_rN_t}.
\end{split}
\end{equation}
Then, the constraint is given \cite{CRLB_kay} by
\begin{equation}\label{47}   
\mathbf{G}=\mathbf{G}I(\mathbf{h}_v)I(\mathbf{h}_v)^\dag.
\end{equation}
Plugging  the (\ref{45}) and (\ref{46}) into (\ref{47}), we obtain $ \mathbf{G}=\mathbf{G}I(\mathbf{h}_v)I(\mathbf{h}_v)^\dag = \mathbf{U_b}$ that holds. This means that the variance of our proposed estimator is finite.
Since the FIM $ I(\mathbf{h}_v) $ in (\ref{45}) is  singular, the expression for the CRLB can be computed following \cite{C_Carbonelli_sparse,CRLB_kay}, which yields,
\begin{equation}\label{48}   
\begin{split}
\mathbf{CRLB_{LSE-SMP}} & \geq \mathbf{C_{LSE-SMP}}= \mathbf{G}I(\mathbf{h}_v)^\dag \mathbf{G}^H \\ 
& =\sigma^{2}_n \left(\left(\mathbf{\bar{S}}\mathbf{U}_{\mathbf{b}}\right)^H\mathbf{\bar{S}}\mathbf{U}_{\mathbf{b}}\right)^{\dag},
\end{split}
\end{equation}
where $\mathbf{C_{LSE-SMP}}$ is the covariance matrix of $\mathbf{\hat{h}}_v^\ast$ for the LSE-SMP estimation \cite{M_Debbah_matrices}. From the CRLB theorem \cite{CRLB_book2}, we know that the unbiased estimator attain the CRLB if and only if
\begin{equation}\label{49_1}   
\frac{ \partial ln \,p(\mathbf{\bar{y}}|\mathbf{h}_v)}{\partial \mathbf{h}_v }=I(\mathbf{h}_v)\left(\mathbf{\hat{h}}_v-\mathbf{h}_v\right),
\end{equation}
always holds.

Plugging (\ref{29}), (\ref{44}) and (\ref{45}) into (\ref{49_1}), this verifies that our proposed LSE-SMP estimators is the MVUE, and can achieve the CRLB under the Assumption 1. 
\end{IEEEproof}

\textbf{\textit{Corollary 1:}} For the deterministic and sparse channel, we have that the MSE of LSE estimator is the upper bound of that of the proposed LSE-SMP estimator. This also be denoted by $\mathbf{MSE_{LSE-SMP}} \leq \mathbf{MSE_{LSE}}$.
\begin{IEEEproof} Firstly, we compute the MSE of the LSE estimator. It can be denoted by
\begin{equation}\label{49_2}   
\begin{split}
\mathbf{MSE_{LSE}} &\!=\! E\{\| \mathbf{\hat{h}}_v-\mathbf{h}_v\|^{2}\} \\
&\!=\!\mathrm{trace}\{ \mathbf{C_{LSE}} \}\!=\!\!\sum\limits_{l= 1}^{N_rN_t}[\mathbf{C_{LSE}}]_{l,l} ,
\end{split}
\end{equation}
where $l \in \{1,2,..,N_rN_t\}$. Similarly, we can obtain the MSE of the LSE-SMP estimator as follows,
\begin{equation}\label{49_3}   
\begin{split}
\mathbf{MSE_{LSE-SMP}} &\!=\! E\{\| \mathbf{\hat{h}}_{v}^*-\mathbf{h}_v\|^{2}\} \\
\!=\! \mathrm{trace}\{ \mathbf{C_{LSE-SMP}} \}&\!=\!\!\sum\limits_{l= 1}^{N_rN_t}[\mathbf{C_{LSE-SMP}}]_{l,l}.
\end{split}
\end{equation}

Since the channel vector is $L$ sparse, $\mathbf{C_{LSE-SMP}}$ has no more than $L$ eigenvalues. Furthermore, we notice that $\mathbf{C_{LSE-SMP}}$  is obtained from the full rank matrix $\mathbf{C_{LSE}}$ by replacing $ N_rN_t-L $ rows and corresponding columns with all zero entries at each index $l $ for which $b_l=0$. Since the $\mathbf{\bar{S}}$ is a non-singular matrix, it is easy to prove that both $\mathbf{C_{LSE}}$ and $\mathbf{C_{LSE-SMP}}$ are the symmetric positive definite matrices, thus their all eigenvalues are greater than zero. We denote $ 0< \lambda_{N_rN_t} \leq \lambda_{N_rN_t-1} \leq ... \leq \lambda_1  $ and  $ 0< \lambda^r_L \leq \lambda^r_{L-1} \leq ... \leq \lambda^r_1 $  as the eigenvalues of $\mathbf{C_{LSE}}$ and $\mathbf{C_{LSE-SMP}}$ respectively. By applying the theorem 4.3.17 in \cite{matrix_Analysis} obtains
\begin{equation}\label{50}   
\lambda_1 \geq \lambda^r_1 \geq \lambda_2 \geq \lambda^r_2 \geq \cdots  \geq  \lambda_L \geq \lambda^r_L \cdots ,
\end{equation}
and therefore
\begin{equation}\label{51}   
\!\mathrm{trace}\{ \mathbf{C_{LSE}} \}\!= \!\!\!\sum\limits_{l= 1}^{N_rN_t} \lambda_l \geq \sum\limits_{l= 1}^{L} \lambda^r_l\! =\!\mathrm{trace}\{ \mathbf{C_{LSE-SMP}} \}.\!\!
\end{equation}

Combining (\ref{51}), and recalling (\ref{49_2}) and (\ref{49_3}), we know that  $\mathbf{MSE_{LSE-SMP}} \leq \mathbf{MSE_{LSE}}$.
\end{IEEEproof}

\subsection{Analysis of Iterative Evolution of LSE-SMP }
In this section, we will analyze the iterative evolution performance of the LSE-SMP. As we mentioned before, it is not easy to analyze the proposed algorithm directly by using the existing methods, i.e., density evolution algorithm \cite{LDPC_Guassion,Maleki_ITD}. Actually, the core of the proposed algorithm is SMP algorithm, and the convergence
behavior of the proposed algorithm is determined by the convergence behavior of the SMP algorithm. Therefore, following the last subsection, we also expect to leverage the alternating minimization method for ruling out the influence of LSE and EM algorithm, and focus on the SMP only. Then, we have the following assumption.

\textbf{Assumption 2:} We have the priori knowledge of non-zero entries of $\mathbf{h}_v$, and its the mean and variance are denoted as $u_h$ and  $\sigma^{2}_h$ respectively.

We design the training signal $\mathbf{\bar{S}}$ that is the Gaussian distribution with the zero mean  and variance $\sigma^{2}_s$. Many previous research results show that  (\ref{32}) can be well approximated by Gaussian densities (see \cite{LDPC_Guassion}). we denote the mean and variance of message $l^{v}_{ij \rightarrow kt}$ at the variable node as $u_v$ and $\sigma^{2}_v$. There is an important condition, called the symmetry condition. For a Gaussian signal with the mean $u_v$ and variance $\sigma^{2}_v$, this condition reduces to $\sigma^{2}_v=2u_v$ for the message updating at variable nodes \cite{LDPC_Richard,LDPC_Guassion,Richardson_01}, which means that we only need to keep the mean $u_v$ of the message $l^{v}_{ij \rightarrow kt}$. Then, we have

\textbf{Theorem 2}: Under the assumption 2, we can obtain a closed-form update rule for variable nodes after any a few iterations. It denotes by
\begin{equation}\label{52}   
u_v(\tau +1)=l_{0}+ \frac{T-1}{\sqrt{4\pi u_v(\tau)}}\int_\mathbb{R} \zeta(l^{v}(\tau))e^{-\frac{(l^{v}(\tau)-u_v(\tau))^2}{4u_v(\tau)}} dl^{v}(\tau),
\end{equation}
with
\begin{equation}   
\begin{split}
\hspace{-3mm}\zeta(l^{v}(\tau)) &= \frac{1}{2}\left(\frac{-\beta-a_1snr^{-1}}{a_2+a_1(snr^{-1}+1)}+\frac{\beta e^{2l^{v}(\tau)}+a_1snr^{-1}}{a_2+a_1snr^{-1}}\right) \\
&\quad\,\,-\mathrm{log} \sqrt{1+\frac{a_1}{a_2+a_1snr^{-1}}} \label{53} ,
\end{split}
\end{equation}
where $k$ denotes the number of iterations, $snr=\frac{\sigma^{2}_s\sigma^{2}_h}{\sigma^{2}_n}$, $\beta= u^{2}_h\sigma^{-2}_h $, $ a_1= (1+e^{l^{v}(\tau)})^{2} $, $ a_2= (N_t-1)e^{l^{v}(\tau)}(1+e^{l^{v}(\tau)}+ \beta) $, and $l^{v}_{ij \rightarrow kt}(\tau) $  is denoted by $ l^{v}(\tau)$.
\begin{IEEEproof}
Proof of the Theorem 2: see the Appendix.
\end{IEEEproof}

Although it is difficult to obtain the exact solution of the mean $ u_v$ and variance $ \sigma^{2}_v$ of $l^{v}(\tau)$, we can study and analyze the performance of SMP algorithm by simulating its Extrinsic Information Transfer (EXIT) chart  \cite{Sten_Exit_Chart,MSE_Chart} based on the Theorem 2. This allows us to take full advantage of properties of the EXIT chart to predict the performance of our proposed algorithm. It is noted that the EXIT chart in this paper is a little difference with the EXIT chart in the paper \cite{Sten_Exit_Chart} and  \cite{MSE_Chart}. The main difference is that we take the variances of variable nodes as the extrinsic information, while the paper \cite{Sten_Exit_Chart} uses the mutual information as the extrinsic information, and the paper \cite{MSE_Chart} uses the MSE as the extrinsic information. In fact, the LLRs of variances message can be transformed into mutual information and MSE directly. Furthermore, we can obtain the Corollary 2 that provides a significant criterion to judge the performance of the estimator.

\textbf{\textit{Corollary 2:}} Assuming that the LSE-SMP has converged, if the variable nodes have the larger convergence variance $ \sigma^{2}_v$, then the LSE-SMP algorithm has the better estimation performance.

\begin{IEEEproof}
Based on the EXIT chart analysis technique, \cite{Sten_Exit_Chart} has proved that there is a connection between the variance $ \sigma^{2}_v$ of the variable node and the bit error rate (BER) after an arbitrary number of iterations.  According to the equation (26) in the \cite{Sten_Exit_Chart}, the estimation of bit error probability can be approximated by the following
\begin{equation}\label{62}   
P_b \approx \frac{1}{2}\mathrm{erfc}\left( \frac{\sigma_v}{2\sqrt{2}} \right)= \frac{1}{2}\mathrm{erfc}\left(\frac{1}{2}\sqrt{u_v}\right),
\end{equation}
where $ \mathrm{erfc}(x)=1-\mathrm{erf}(x)=\frac{2}{\sqrt{\pi}} \int_x^{\infty} e^{-t^2} dt$ is the complementary error function, and is a monotonically decreasing and continuous on $(-\infty, +\infty)$. This means that the BER will decrease with the increase of the convergence variance $ \sigma^{2}_v$. Therefore, the Corollary 2 is proved.
\end{IEEEproof}

\textbf{\textit{Remark 4:}} The Theorem 2 shows the basic relationship between $u_v(\tau +1)$, $u_v(\tau)$ and some parameters, i.e., training sequence length $T$, $snr$, coefficient of variation of the channel $\beta$, etc. Moreover, the Corollary 2 provides a remarkable connection between the variance $ \sigma^{2}_v$ of the variable node and the performance of the algorithm, which provides a significant criterion for our proposed algorithm.  Based on the Theorem 2 and Corollary 2, we take the advantage of the EXIT chart analysis to obtain a few insights on the proposed algorithm in the numerical results section. 
\begin{figure} \vspace{-5.5mm}
 \begin{center}
  \includegraphics[width=74mm]{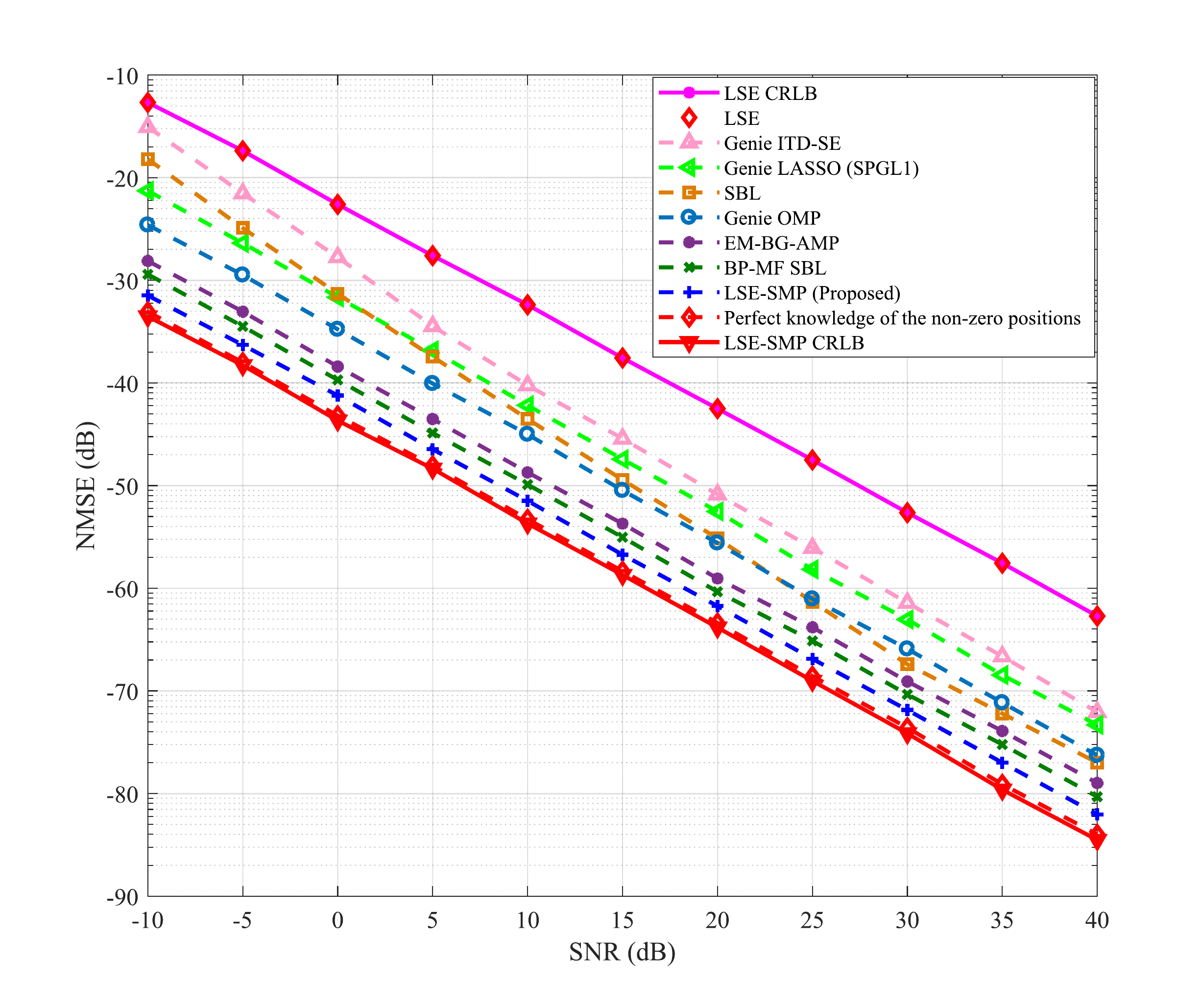}\\ \vspace{-4.5mm} 
  \caption{This figure shows the SNR versus the average NMSE performance of various estimate algorithms. In the figure, $N_t=32, N_r=64, \eta= 0.007$, $\beta =10$ with the $\sigma^{2}_h=10$, and the training sequence length $T=64$.
  The result shows the proposed LSE-SMP estimator exhibits the best NMSE performance among the tested algorithms.}
  \label{fig:Comparison} \vspace{-7mm}
 \end{center}
\end{figure}

\section{NUMERICAL RESULTS}
In this section, we report the results of a detailed numerical study on the performance of our proposed LSE-SMP algorithm using the Monte-Carlo simulations.  
For all numerical study, we considered the channel estimation problem in a $ 32 \times 64 $ mmWave MIMO system. The channel vector $\textbf{h}_v$ was randomly generated based on (3) and (8), and the non-zero entries follow a Gaussian distribution. Throughout, we considered SNR $\triangleq E\{\|\mathbf{\bar{S}} \|^{2}_F / \|\mathbf{\bar{n}} \|^{2}_F\}$ in the interval $[-10, 40]dB$; defined the coefficients of variation of $ \mathbf{h}_v $ as $ \beta =u^{2}_h/\sigma^{2}_h$ with the $\sigma^{2}_h=10$; and the performance metric was the  Normalized Mean Square Error (NMSE), given by $ E\{\frac{ \| \mathbf{\hat{h}}_{v}^*-\mathbf{h}_{v}\|^{2}_F}{\| \mathbf{h}_{v}\| ^{2}_F}\}$. There are 500 different channel realizations and the average results are reported.  \vspace{-0mm}

\subsection{Performance Comparison}
Fig. \ref{fig:Comparison} shows the average channel estimation NMSE performance of the proposed LSE-SMP algorithm, LSE, Genie aided ITD-SE \cite{C_Carbonelli_sparse}, genie-tuned LASSO (via SPGL1 \cite{LASSO_test}), SBL \cite{SBL} (via the T-MSBL \cite{T_SBL}), genie-tuned OMP \cite{OMP01}, EM-BG-AMP \cite{P_Schniter_EM} (in sparse mode) and BP-MF SBL\cite{Guo_Qing}. All algorithms were run under the suggested defaults to obtain their best performance by the varied maximum number of iterations.
Additionally, we also compute the CRLB for the classical LSE and the proposed LSE-SMP estimator. The result shows that our proposed LSE-SMP estimator exhibits the best NMSE performance among the tested algorithms, and reduces the NMSE by $3.5dB$ relative to EM-BG-AMP and $1.5dB$ relative to BP-MF SBL. As expected, the CRLB for the proposed LSE-SMP is the lowest, and this result is consistent with that of classical LSE estimator with the perfect knowledge of the non-zero positions. It is also seen from the Fig. \ref{fig:Comparison} that the gap between LSE-SMP and LSE-SMP CRLB is much smaller, about $1.8dB$. This gap is partly due to the errors in the detection of the non-zero positions in sparse message passing phase and partly to the fact that all our detection strategies rely on a coarse initial estimate of the channel.
\begin{figure} \vspace{-4mm}
  \begin{center}
  \includegraphics[width=74mm]{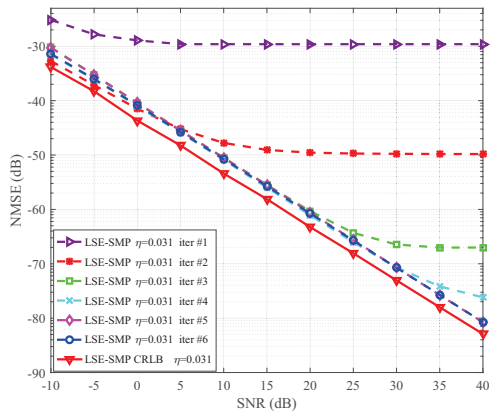} \vspace{-4mm} \\  
  \caption{This figure shows the average NMSE performance of the LSE-SMP estimator and its CRLB versus SNR under different turbo iterations. In the figure, $N_t=32, N_r=64, \eta= 0.031$, $\beta =10$ with the $\sigma^{2}_h=10$, and the training sequence length $T=64$. This result shows that the LSE-SMP algorithm reaches the convergence just need five iterations. }
  \label{fig:iterations} \vspace{-6mm}
  \end{center}
\end{figure}
\begin{figure}\vspace{-1mm}
  \begin{center}
  \includegraphics[width=74mm]{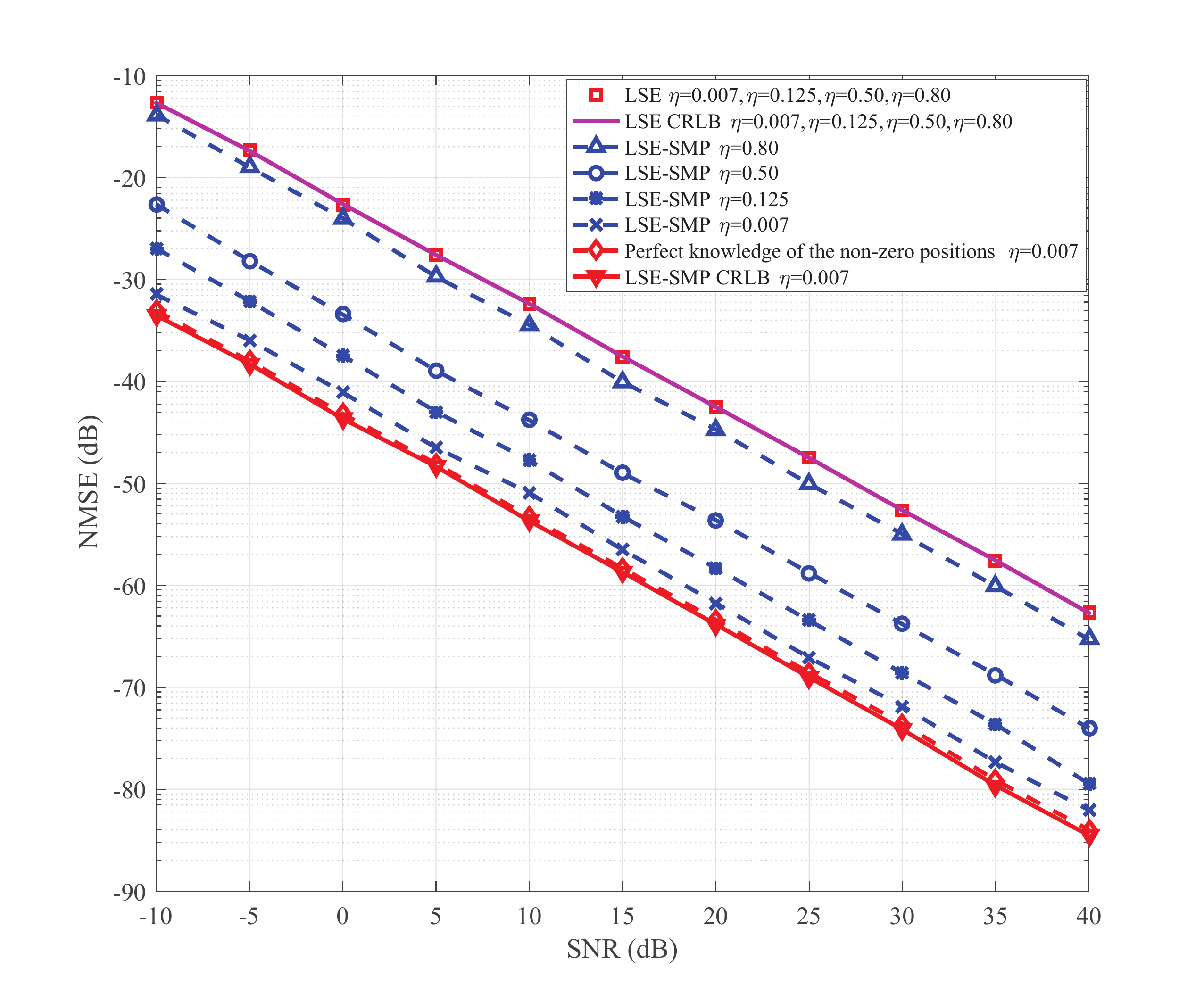} \vspace{-4mm}\\  
  \caption{ This figure shows SNR versus the average NMSE performance comparison of LSE-SMP, LSE channel estimates and their CRLBs for different sparsity ratios. In the figure, $N_t=32, N_r=64, \eta \in \{ 0.007, 0.125 ,0.50 ,0.80\} $, $T=64$, $\beta =10$ with the $\sigma^{2}_h=10$, and the iterations $=6$. It can be seen that the LSE-SMP CRLB is the lower bound of the proposed LSE-SMP and the LSE CRLB is the upper bound.}
  \label{fig:different_sparsity} \vspace{-6mm}
  \end{center}
\end{figure}


\subsection{Effect of Iterations }
Fig. \ref{fig:iterations} shows the average channel estimation NMSE performance for the LSE-SMP algorithm under several turbo iterations.
The result shows that the NMSE performance of the LSE-SMP algorithm will be lower with the increasing of iterations, and we also find that the gap of the NMSE performance between the adjacent iterations for the LSE-SMP algorithm will be smaller with more iterations. The main reason is that the parameters of sparsity ratio $\eta$ and non-zeroes position vector $\mathbf{b}$ are estimated more and more accurate. On the other hand, after the fifth turbo iteration, the NMSE performance have no significant improvement and it is very close to our analyzed LSE-SMP CRLB. This demonstrates that the convergence speed of the LSE-SMP algorithm is fast (just need five iterations).
\subsection{Effect of Sparsity Ratios}
For further investigating the effect of sparsity ratio to our proposed algorithm, we ran the algorithm to obtain its best performance and changed the sparsity ratio $\eta$ from $0.007$ to $ 0.80$. Simulation results in Fig. \ref{fig:different_sparsity} show that CRLB of the LSE-SMP is the lower bound and CRLB of the LSE is the upper bound. When the channel become more sparse, the performance of the proposed LSE-SMP will be better and approaches the lower bound. These results verified the analysis of  Theorem 1 and Corollary 1.  In addition, these results also show that the LSE-SMP is able to exploit the sparsity of the channel. To be specific, The NMSE performance of the LSE keeps unchanged under different sparsity ratios, while the NMSE of the LSE-SMP will decrease with the decrease of sparsity ratios. This indicates that the LSE-SMP scheme will be very suitable for channel estimation for mmwave systems since the channel of mmwave systems is sparse.
\begin{figure*} \vspace{-8mm}
  \begin{center}
  \includegraphics[width=138mm,height=84mm,keepaspectratio]{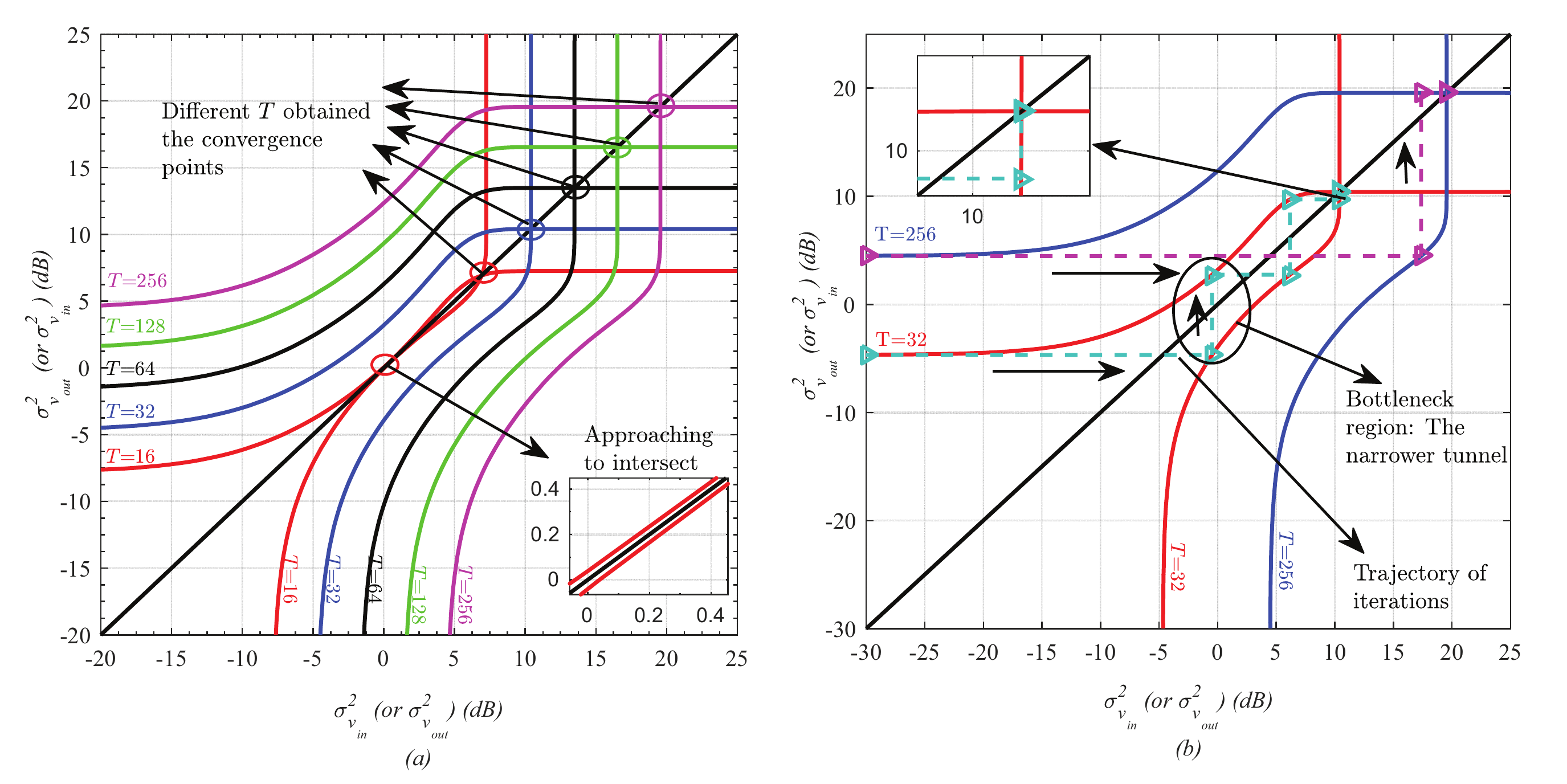} \vspace{-4mm}\\  
  \caption{ This figure shows the EXIT chart analysis of the LSE-SMP algorithm, considering the case that $N_t=32, N_r=64, SNR=10 dB, \beta =10 $ with $ \sigma_{h}^2=10 $, and $ \eta= 0.125$. $ \sigma^2_{in}$  and  $ \sigma^2_{out} $ are the input and output variance messages of the variable nodes in the LLRs form. (a), $ T\in \{ 16, 32, 64, 128, 256\}$ (b), $ T\in \{32, 256\}$.  }
  \label{fig:Training_combine} \vspace{-8mm}
  \end{center}
\end{figure*}

\subsection{Effect of Training Sequence Length }
In Fig. \ref{fig:Training_combine}(a) and (b), we investigate the effect of different training sequence lengths by tracking the input and output LLRs of variance messages ($ \sigma^2_{in}$  and  $ \sigma^2_{out} $) of the variable nodes of the LSE-SMP algorithm.
From the results of Theorem 2 and Corollary 2, we know that the performance of the LSE-SMP algorithm can be measured by the convergence variances of the variable nodes. Therefore, we can leverage the EXIT chart based technique to analyse the system performance and design optimal parameters for some specific application scenarios.  More specifically. Fig. \ref{fig:Training_combine}(a) shows that the variance messages always converges to only one fixed point, and the longer the training sequence length, the better the estimation performance (BER). On the other hand, Fig. \ref{fig:Training_combine}(b) shows that the increase in training sequence length will reduce the number of iteration needed, but at the cost of the overall computational time as the computational complexity of each iteration is exponential to the demission of matrix \cite{Loeliger_factor_graph,factor_graph_LDPC,P_Schniter_facror_graph,Lei2015_TWC}. Therefore, through the EXIT chart analysis, we can optimize the training sequence length to achieve a balance between the estimation performance and computational time. In addition, we also find that the shorter training sequence will result in the narrower space between the input and output variance traces. With the decrease of training sequence length, two traces of  $ \sigma^2_{in} $ and $ \sigma^2_{out} $ will approach to intersect. If two traces have more one intersections, this means that the proposed algorithm fails to converge. To be specific, $T=16$ is the shortest training sequence that can make the proposed LSE-SMP work and converge under the setting of figures. Similarly, another an important function of the EXIT chart analysis is to predict the training sequence length under given the BER according to the Theorem 2 and Corollary 2.

\subsection{Effect of the Coefficient of Variation of the Channel }
\begin{figure} \vspace{-2mm}
  \begin{center}
  \includegraphics[width=75mm]{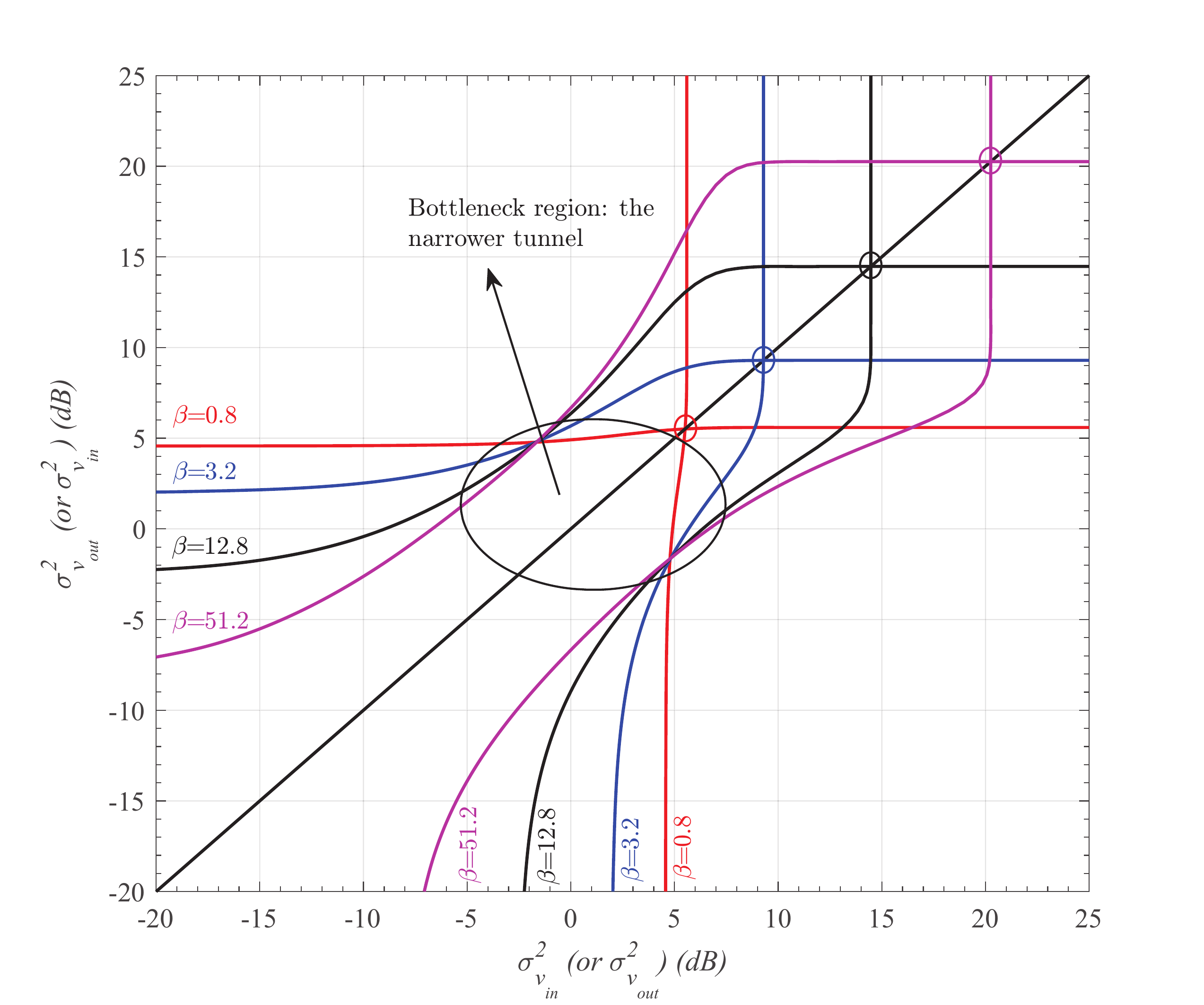} \vspace{-2mm}\\  
  \caption{This figure shows the EXIT chart analysis of the LSE-SMP algorithm with $T= 64$, $N_t=32, N_r=64 $ and $ SNR= 10dB$, where we set the coefficients of variation $\beta \in \{0.8, 3.2, 12.8, 51.2\}$ under the $ \sigma_{h}^2=10 $. The result shows that the more dispersion channel condition will leads to the lower system performance, while needs the less time to achieve the convergence. }
  \label{fig:alpha} \vspace{-6mm}
  \end{center}
\end{figure}
Fig. \ref{fig:alpha} presents the effect of the coefficient of variation $ \beta $ of the channel vector $\mathbf{h}_v$ for the proposed LSE-SMP algorithm.
Then, we plotted the EXIT chart for the coefficient of variation. We can see that there is only one convergent variance point for the different coefficients of variation,  and the convergent variance point will be higher with the increase of the coefficient of variation, while the \textsl{bottleneck region} that denotes the narrower region between the $ \sigma^2_{in}$  and  $ \sigma^2_{out} $  traces keeps unchanged. Therefore, this also means that the higher variance point needs more iterations to achieve. In other words, these results indicate that the more dispersion channel condition will leads to the lower system performance, but needs the less time to achieve the convergence. 

\subsection{ Complexity of the Algorithms }
\begin{figure} \vspace{-2mm}
  \begin{center}
  \includegraphics[width=77mm]{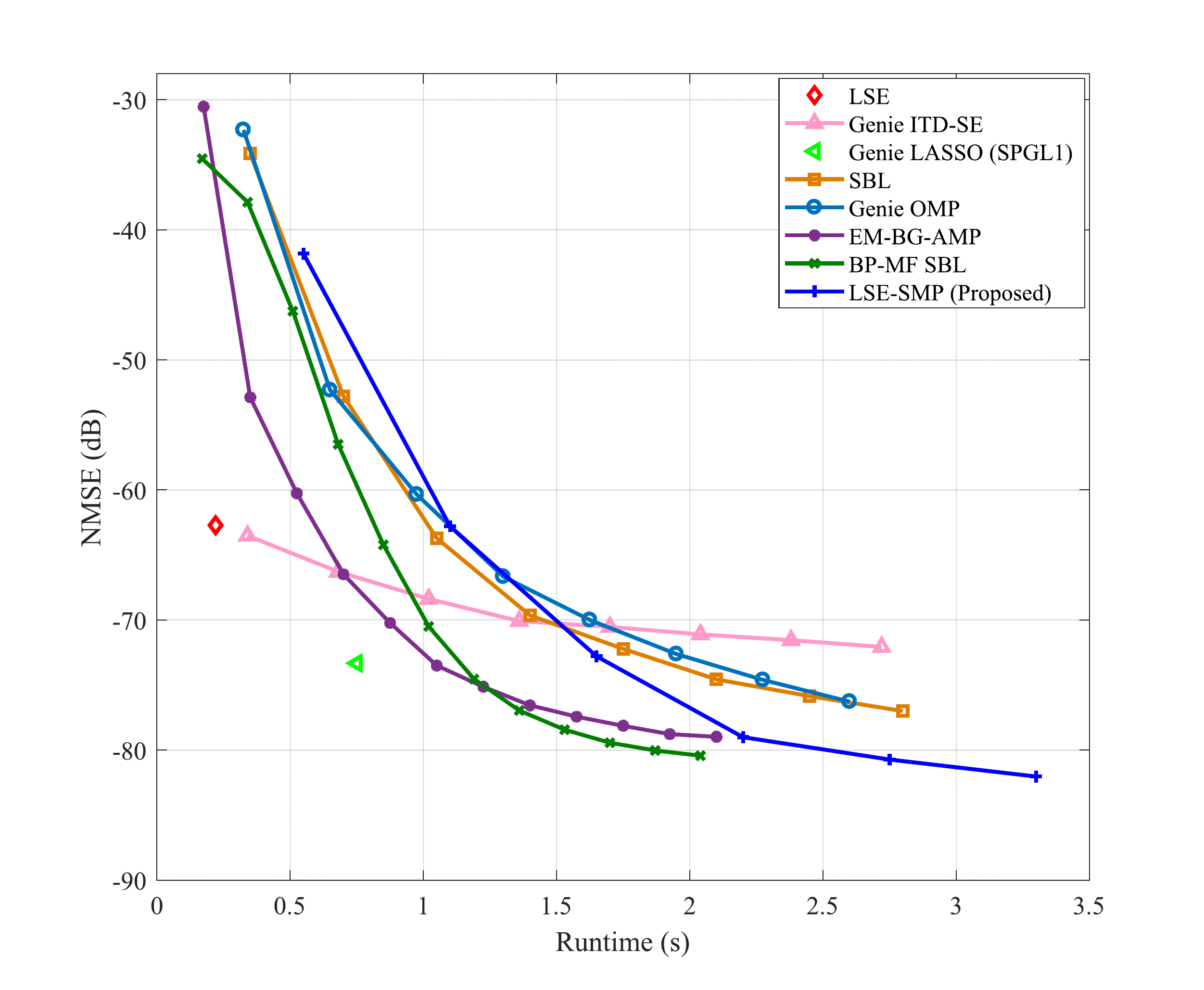} \vspace{-3mm}\\  
  \caption{  This figure shows the NMSE versus the runtime of various algorithms. In the figure, $N_t=32, N_r=64, SNR=40 dB, \beta =10 $ with $ \sigma_{h}^2=10 $, $ \eta= 0.007,$ and $ T=64.$  This result indicates that the proposed LSE-SMP obtains the best performance-complexity trade-off than the tested algorithms when the NMSE is less than $-80dB$.  }
  \label{fig:complexity} \vspace{-6mm}
  \end{center}
\end{figure}

Fig. \ref{fig:complexity} shows the NMSE versus the runtime of various algorithms. We evaluate the runtime of each algorithm on a typical personal computer.
By varying the maximum number of iterations in ITD-SE, SBL, OMP, EM-BG-AMP, BP-MF SBL and LSE-SMP algorithms, we obtained their NMSE-runtime frontier. The other two algorithms are represented by two points. We notice that the ITD-SE obtains the best performance-complexity trade-off when the NMSE is larger than $-66dB$, LASSO gives the best trade-off when the NMSE is between $-66dB$ and $-73dB$, EM-BG-AMP gives the best trade-off when the NMSE is between $-73dB$ and $-76dB$, BP-MF SBL gives the best trade-off when the NMSE is between $-76dB$ and $-80dB$, and LSE-SMP is the best when the NMSE is less than $-80dB$. In other words, the proposed LSE-SMP algorithm obtains the best performance than other tested algorithms, although it may need to spend more time.

\section{CONCLUSION}
We have presented a sparse channel estimation algorithm (LSE-SMP) for mmWave MIMO systems, which leverages both virtues of the SMP and LSE algorithms. We have analyzed the CRLB of the proposed LSE-SMP algorithm, and showed that the algorithm is MVUE under the assumption that we have the partial priori knowledge of the channel. Next, we have also shown an EXIT chart-based analysis technique for the convergence analysis of the key part of the proposed algorithm, and for the selection of optimal design parameters. Simulation experiments have verified that the proposed algorithm reduces the NMSE by about $2dB$ relative to the best of existing algorithms. In addition, it has been shown that the proposed algorithm typically needed only five turbo iterations to approximately achieve CRLB. Since the correlation among the adjacent entries in the beam domain channel was ignored in this paper, a possible research direction is to take it into consideration in the future work.

\appendix
Then, (\ref{32}) simply becomes
\begin{figure*} \vspace{-3mm}
\begin{equation}\label{54}   
\begin{split}
\hspace{-3mm} E\{l^{v}_{ij \rightarrow kt}(\tau +1)\} &=u_v(\tau +1)=E \left\{l_{0}+\sum \limits_{\kappa \neq k}^{N_s}l^{s}_{\kappa t\rightarrow ij}(\tau)\right\} \\
&=E\left\{l_0+\sum \limits_{\kappa \neq k}^{N_s} \mathrm{log} \frac{ \mathcal{N}(y_{\kappa t}|e^{s}_{\kappa t\rightarrow ij}(\tau)+s_{\kappa t,ij}\hat{h}_{ij}(\tau),v^{s}_{\kappa t\rightarrow ij}(\tau)+s^{2}_{\kappa t,ij}v_{h_{ij}}(\tau))}{\mathcal{N}(y_{\kappa t}|e^{s}_{\kappa t\rightarrow ij}(\tau),v^{s}(\tau)_{\kappa t\rightarrow ij}(\tau))} \right\}. \\
\end{split}
\end{equation} \vspace{-3mm}
\end{figure*}
Plugging (\ref{22}) into (\ref{54}), and we obtain the update as following,
\begin{equation}\label{55}   
\begin{split}
\hspace{-1mm} u_v(&\tau +1) 
=l_{0}+E\bigg \{ \sum \limits_{\kappa \neq k}^{N_s} \frac{(y_{\kappa t}-e^{s}_{\kappa t\rightarrow ij}(\tau)-s_{\kappa t,ij}\hat{h}_{ij}(\tau))^2} {-2(v^{s}_{\kappa t\rightarrow ij}(\tau)+s^{2}_{\kappa t,ij}v_{h_{ij}}(\tau))} \\
& + \frac{(y_{\kappa t}-e^{s}_{\kappa t\rightarrow ij}(\tau))^2} {2(v^{s}_{\kappa t\rightarrow ij}(\tau))} - \mathrm{log } \sqrt{\frac{v^{s}_{\kappa t\rightarrow ij}(\tau)+s^{2}_{\kappa t,ij}v_{h_{ij}}(\tau)}{v^{s}_{\kappa t\rightarrow ij}(\tau)}} \bigg\}.\\
\end{split}
\end{equation}

In order to be convenient for analysis, we define equation (\ref{56}) as follow,
\begin{figure*} \vspace{-2mm}
\begin{equation}\label{56}   
\zeta(l^{v}_{ij \rightarrow kt}(\tau))= \frac{(y_{kt}-e^{s}_{kt\rightarrow ij}(\tau))^2} {2(v^{s}_{kt\rightarrow ij}(\tau))}-\frac{(y_{kt}-e^{s}_{kt\rightarrow ij}(\tau)-s_{kt,ij}\hat{h}_{ij}(\tau))^2} {2(v^{s}_{kt\rightarrow ij}(\tau)+s^{2}_{kt ij}v_{h_{ij}}(\tau))} -\mathrm{log } \sqrt{\frac{v^{s}_{kt\rightarrow ij}(\tau)+s^{2}_{kt,ij}v_{h_{ij}}(\tau)}{v^{s}_{kt\rightarrow ij}(\tau)}}.
\end{equation} \vspace{-1mm}
\end{figure*}

Since transmit and receive antennas are independent and symmetric, we omit the subscript of $l^{v}_{ij \rightarrow kt}(\tau) $, and use $ l^{v}(\tau)$ to denote the $l^{v}_{ij \rightarrow kt}(\tau) $ in the following paper. Due to $s_{kt,ij} \sim \mathcal{N}(0,\sigma^{2}_s) $, we can obtain $ E(s^{2}_{kt,ij})=Var(s_{kt,ij})+E(s_{kt,ij})^2=\sigma^{2}_s$. Similarly, we can get $ E(n^{2}_{kt})=Var(n_{kt})+E(n_{kt})^2=\sigma^{2}_n$. Plugging (\ref{21}), (\ref{21_1}) and (\ref{31_2}) into the above expression, we obtain
\begin{equation}\label{58}   
\begin{split}
\hspace{-0mm} \zeta(l^{v}(\tau))&= -\mathrm{log} \left( \sqrt{\frac{ \frac{ (N_t-1)\sigma^{2}_s} {1+e^{-l^{v}(\tau)}}(\sigma^{2}_h+\frac{u^{2}_h}{1+e^{l^{v}(\tau)}})+\sigma^{2}_s\sigma^{2}_h+\sigma^{2}_n}{\frac{(N_t-1) \sigma^{2}_s}{1+e^{-l^{v}(\tau)}}(\sigma^{2}_h+\frac{u^{2}_h}{1+e^{l^{v}(\tau)}})+\sigma^{2}_n} } \right) \\
& + \frac{ \frac{-\sigma^{2}_su^{2}_h}{(1+e^{l^{v}})^2} -\sigma^{2}_n }{2 \left( \frac{ (N_t-1)\sigma^{2}_s} {1+e^{-l^{v}(\tau)}}(\sigma^{2}_h+\frac{u^{2}_h}{1+e^{l^{v}(\tau)}})+\sigma^{2}_s\sigma^{2}_h+\sigma^{2}_n \right)} \\
& +\frac{ \frac{\sigma^{2}_su^{2}_h}{(1+e^{-l^{v}(\tau)})^2} + \sigma^{2}_n }{2 \left( \frac{ (N_t-1)\sigma^{2}_s} {1+e^{-l^{v}(\tau)}}(\sigma^{2}_h+\frac{u^{2}_h}{1+e^{l^{v}(\tau)}})+\sigma^{2}_n \right)}. \\
\end{split}
\end{equation}

To simplify the expression, we yield equation (\ref{59}).
\begin{figure*}   \vspace{-3mm}
\begin{equation}\label{59}   
\begin{array}{l}  \vspace{-0mm}
\hspace{-3mm}\zeta(l^{v}(\tau))= -\mathrm{log} \left( \sqrt{1+\frac{(1+e^{l^{v}(\tau)})^{2}}{(N_t-1)e^{l^{v}(\tau)}(1+e^{l^{v}(\tau)}+u^{2}_h\sigma^{-2}_h)+(1+e^{l^{v}(\tau)})^2snr^{-1}}} \right) + \frac{1}{2}
\\\hspace{-3mm}\left(\frac{-u^{2}_h\sigma^{-2}_h-(1+e^{l^{v}(\tau)})^{2}snr^{-1}}{(N_t-1)e^{l^{v}(\tau)}(1+e^{l^{v}(\tau)}+u^{2}_h\sigma^{-2}_h)+(1+e^{l^{v}(\tau)})^2(snr^{-1}+1)}+
\frac{u^{2}_h\sigma^{-2}_he^{2l^{v}(\tau)}+(1+e^{l^{v}(\tau)})^{2}snr^{-1}}{(N_t-1)e^{l^{v}(\tau)}(1+e^{l^{v}(\tau)}+u^{2}_h\sigma^{-2}_h)+(1+e^{l^{v}(\tau)})^2snr^{-1}}\right),
\end{array}       \vspace{-0mm}
\end{equation} \vspace{-8mm}
\end{figure*} \\
where $snr=\frac{\sigma^{2}_s\sigma^{2}_h}{\sigma^{2}_n}$.

The above expression looks like complex, however, we find that it is symmetric and has a few common terms. Then, we can define  $\beta= u^{2}_h\sigma^{-2}_h $, $a_1= (1+e^{l^{v}(\tau)})^{2} $, and $ a_2= (N_t-1)e^{l^{v}(\tau)}(1+e^{l^{v}(\tau)}+ \beta) $. The $ \phi(l^{v}(\tau)) $ can be denoted by
\begin{equation}\label{60}   
\begin{split}
\zeta(l^{v}(\tau)) &= \frac{1}{2}\left(\frac{-\beta-a_1snr^{-1}}{a_2+a_1(snr^{-1}+1)}+\frac{\beta e^{2l^{v}(\tau)}+a_1snr^{-1}}{a_2+a_1snr^{-1}}\right) \\
&\quad\,\,-\mathrm{log} \sqrt{1+\frac{a_1}{a_2+a_1snr^{-1}}} ,
\end{split}
\end{equation}

Since $l^{v} $ is Gaussian with the mean $ u_v$ and variance $ 2u_v$, we obtain the following update expression by the definition of the expectation,
\begin{equation}\label{61}   
u_v(\tau +1)=l_{0}+ \frac{T-1}{\sqrt{4\pi u_v(\tau)}}\int_\mathbb{R} \zeta(l^{v}(\tau))e^{-\frac{(l^{v}(\tau)-u_v(\tau))^2}{4u_v(\tau)}} dl^{v}(\tau).
\end{equation}
Therefore, we have the Theorem 2.

\section*{Acknowledgement}
The work of C. Yuen was supported by the MIT-SUTD International design center and NSFC 61750110529 Grant, and that of C. Huang by the PHC Merlion PhD program. The authors would like to thank the associate editor and the reviewers for their valuable comments and suggestions.

\end{document}